\documentclass[a4paper,12pt]{article}
\usepackage{amsmath}
\usepackage{amsthm}
\usepackage{amsfonts}
\usepackage{mathrsfs}
\usepackage{graphicx}
\usepackage{hyperref}
\usepackage{float}
\usepackage{subcaption}
\usepackage{multirow, enumitem}
\usepackage[sorting=none]{biblatex}
\bibliography{references}

\numberwithin{equation}{section}

\hypersetup{ pdftitle={Report 2},
pdfauthor={Fiki T. Akbar}, 
pdfkeywords={Scalar Wave Equation, Non minimal Coupling, Existence}, 
bookmarksnumbered, pdfstartview={FitH}, urlcolor=blue, }

\begin{document}
\pagestyle{plain}




\title{\LARGE\textbf{Isotropic and Anisotropic Radiating Gravastars with Various Matter Types of Thin Shell and Interior}}

\author{{Hasan Al-Asy'ari, Fiki Taufik Akbar{\footnote{Corresponding Author}}, Bobby Eka Gunara}\\ \\
\textit{\small Theoretical Physics Laboratory}\\
\textit{\small Theoretical High Energy Physics Research Division,}\\
\textit{\small Faculty of Mathematics and Natural Sciences,}\\
\textit{\small Institut Teknologi Bandung}\\
\textit{\small Jl. Ganesha no. 10 Bandung, Indonesia, 40132}\\ \\
\small email: hasan.alasyari0710@gmail.com, ftakbar@itb.ac.id, bobby@itb.ac.id}

\date{}

\maketitle




\begin{abstract}
	In this paper, we investigate models of radiating gravastars with both isotropic and anisotropic interiors, incorporating various types of thin shell matter. For the isotropic interior case, we consider a thin spherical shell characterized by an equation of state in which its pressure is proportional to its mass density, enclosing a de Sitter spacetime and surrounded by Vaidya exterior spacetime. Our analysis reveals that stable gravastars can form under specific scenarios of radiative mechanisms and for certain thin shell matter types. In addition, we also show and discuss in brief the possibility of existence of stable radiating anti-de Sitter gravastar formation. For the anisotropic interior, we use an anisotropic dark energy model with a Tolman-Matese-Whitman (TMW) mass function. We explore several thin shell matter types: standard, dark energy, and repulsive phantom. Our findings indicate that stable gravastars can also emerge in this context, particularly with standard and repulsive phantom thin shells. Furthermore, our results suggest that the density of black holes is consistently higher than that of gravastars and normal stars, regardless of the type of matter in the thin shell. This observation supports the notion that gravastars and black holes are distinct entities, reinforcing the theoretical distinction between these two types of compact objects.

\end{abstract}




\section{Introduction}

Einstein's theory of general relativity predicts the existence of black holes, which are extraordinary cosmic objects characterized by a singularity at their core and an event horizon at their boundary. This event horizon prevents any information from escaping, rendering the interior of black holes inaccessible to external observers. As a result, black holes present a profound challenge to our understanding of the universe, particularly due to the singularity's nature, which defies current physical theories. To address these challenges, there are some attempts to propose alternative models, one of which is the gravastar (gravitational vacuum star).

The gravastar concept was introduced by P. O. Mazur and E. Mottola in 2002 \cite{mottola2002gravitational, mazur2004gravitational, mazur2004dark}. They proposed that, instead of collapsing into a singularity, a star undergoing gravitational collapse might undergo a phase transition near the event horizon. This transition results in the formation of a de Sitter spacetime characterized by negative pressure ($p = \rho < 0$) within the object where $\rho$ is the mass density, counteracting gravitational collapse. The gravastar model comprises a de Sitter interior, a thin shell of stiff matter ($p = \rho$), and an exterior region described by Schwarzschild spacetime ($p = \rho = 0$). Despite the theoretical differences, from an external viewpoint, a gravastar appears indistinguishable from a black hole, offering a novel approach to overcoming the information loss paradox associated with black holes.

Visser and Wiltshire (VW) further explored gravastar stability using a three-layer model based on the potential formulation of the thin shell's equation of motion \cite{visser2004stable},

\begin{equation}
	\frac{1}{2}\left(\frac{dR}{d\tau}\right)^2+V(R)=0\;,
	\label{1}
\end{equation}
where $\tau$ denotes the proper time.
They identified two types of stable gravastars: the static gravastar, which satisfies the conditions
\begin{equation}
    V(R_0) = 0,\qquad \frac{dV}{dR}(R_0) = 0,\qquad \frac{d^2V}{dR^2}(R_0) > 0,
\end{equation}
where \( R_0 \) is the radius at which these conditions are fulfilled, and the "bounded excursion" gravastar, where the shell oscillates between two radii, \( R_1 \) and \( R_2 \), with
\begin{equation}
    V(R_1) = 0,\qquad \frac{d^2V}{dR^2}(R_1) \le 0,\qquad V(R_2) = 0,\qquad \frac{d^2V}{dR^2}(R_2) \ge 0,
\end{equation}
and \( V(R) < 0 \) for \( R \in (R_1, R_2) \).

Understanding such objects necessitates investigating their interaction with surrounding matter. One significant aspect of this is radiation. While substantial research has been conducted on black hole radiation, known as Hawking radiation \cite{hawking1974black}, there has been limited exploration of radiation from gravastars. R. Chan et al. initiated this exploration with their radiating gravastars model, combining a de-Sitter interior with a Vaidya exterior \cite{chan2011radiating}. Subsequent studies by Nakao et al. examined radiation power and phase time during gravastar formation, introducing the OHN (Okabayashi, Harada, and Nakao) model \cite{nakao2022quantum, okabayashi2022robustness}. Further work by R. Chan et al. has explored the stability of gravastar models with anisotropic dark energy interiors and Schwarzschild exteriors \cite{chan2009stable, chan2011gravastars}, which differ in the mass distribution within the interior. These studies build on earlier findings by Cattoen et al., who demonstrated the necessity of anisotropic pressure in gravastars \cite{cattoen2005gravastars} and they were supported by DeBenedictus et al.'s proposal of several anisotropic gravastars model \cite{debenedictis2006gravastar}.


The findings of Cattoen et al. inspired researchers for further investigation of anisotropic gravastars in many cases and contexts. Lobo and Arellano investigated anisotropic gravastars with nonlinear electrodynamics \cite{lobo2007gravastars}. Stelea et al. studied magnetized gravastars with anisotropic constant energy density interior proposed by Lobo \cite{lobo2006stable} and they find that the presence of magnetic field in such gravastars affects anisotropic matter distribution in the shell \cite{stelea2018magnetized}. Amat et al. studied anisotropic gravastar in the  $f(R,T)$ gravity framework \cite{azmat2022study}. 
Other researchers investigated gravastars in the braneworld gravity framework and found that anisotropy is an essential feature of gravastars in that framework \cite{arbanil2019gravastar,sengupta2020gravastar,ray2022gravastar}.
Sakti and Sulaksono in their study about dark energy stars with phantom field also demonstrated that the dark energy stars can be gravastars which have positive anisotropy factor \cite{sakti2021dark}. Jampolski and Rezolla revisited gravastars model with anisotropic pressure and demonstrated the possibilty of finding nested anisotropic gravastars which they call "nestar" \cite{jampolski2024nested}.

In this paper, we extend the analysis of radiating gravastars by modifying the model proposed by R. Chan et al. \cite{chan2011radiating}. Our study comprises two main modifications. This first modification generalizes previous studies on stable gravastars with a three-layer model \cite{rocha2008bounded,rocha2008stable,chan2011radiating} and investigate variations in the thin shell while maintaining an isotropic de Sitter spacetime interior. We formulated the potential, its first and second derivatives, and the mass as functions of $\eta$ and analyzed the possibility of a stable gravastar. Here, we also expand our investigation into anti-de Sitter interior case and present in short several parameters in our model which can produce radiating anti-de Sitter gravastars. The stability of anti-de Sitter gravastar interior has been briefly explored by Visser et al. \cite{visser2004stable} and studied further with not only Schwarzschild exterior but also Reissner–Nordstrom exterior by Carter \cite{carter2005stable}.

Second, we examine cases with both modified thin shells and anisotropic dark energy interiors, represented by the Tolman-Matese-Whitman (TMW) mass function. We demonstrated several stable radiating gravastars with an anisotropic dark energy fluid interior and TMW mass function. We identified three possible interior types: standard, dark, and repulsive phantom, but not attractive phantom.

The paper is organized as follows: Section \ref{sec:isotropic} explores gravastars with isotropic interiors and various thin shell matter types. Section \ref{sec:anisotropic} addresses gravastars with anisotropic dark energy interiors and diverse thin shell matters. Finally, Section \ref{sec:conclusion} summarizes our findings and conclusions.


\section{Isotropic Interior}
\label{sec:isotropic}

\subsection{Gravastar Model for Isotropic Interior Case }
In this section, we explore the three-layer gravastar model as studied by Visser and Wiltshire \cite{visser2004stable}. The interior spacetime of the gravastar is given by de-Sitter metric, 
\begin{equation}
	ds_i^2=-f_1dt^2+f_2dr^2+r^2(d\theta^2+\sin^2\theta d\phi^2),
	\label{4}
\end{equation}
where $f_1=1-(r/L)^2$, $f_2= \left[1-(r/L)^2\right]^{-1} = f_1^{-1}$, $L=\sqrt{3/\Lambda}$, and $r$ is the radial coordinate. Since we consider a radiating gravastar, then the exterior spacetime described by Vaidya metric,
\begin{equation}
	ds_e^2=-Fdv^2-2d\bar{r}dv+\bar{r}^2(d\theta^2+\sin^2\theta d\phi^2),
	\label{5}
\end{equation}
where $F=1 - 2m(v)/\bar{r}$. 

Let us then consider he hypersurface metric given by
\begin{equation}
	ds_\Sigma^2=-d\tau^2+R^2(\tau)(d\theta^2+\sin^2\theta d\phi^2),
	\label{6}
\end{equation}
where $\tau$ is the proper time. Since the metric on the hypersurface should be equal, $ds_i^2=ds_e^2=d_\Sigma^2$, then we obtain the relation $r_\Sigma=\bar{r}_\Sigma=R$, and
\begin{equation}
	f_1\dot{t}^{2}-f_2\dot{R}^{2}=1,
	\label{7}
\end{equation} 
\begin{equation}
	\bigg[F+\frac{2\dot{R}}{\dot{v}}\bigg]\dot{v}^{2}=1,
	\label{8}
\end{equation}
where the dot denotes differentiation with respect to the proper time. The interior and exterior normal vectors to the thin shell are:
\begin{equation}
	\begin{aligned}
		n_{\alpha}^i &=(-\dot{R},\dot{t},0,0),\\
		n_{\alpha}^e &=(-\dot{R},\dot{v},0,0)\;,
	\end{aligned}
\end{equation}
respectively.

The mass of the shell can be written as \cite{lake1979thin}

\begin{equation}
	M=K_{\theta\theta}^{i}-K_{\theta\theta}^{e} = \dot{t}\Bigg[1-\bigg(\frac{R}{L}\bigg)^{2}\Bigg]+\dot{v}(2m-R) -R\dot{R},
	\label{10}
\end{equation}
where $K_{\theta\theta}^{i}$ and $K_{\theta\theta}^{e}$ denote the extrinsic curvature of the interior and exterior, respectively. Substituting \eqref{7} and \eqref{8} to \eqref{10}, the mass of the shell can be rewritten as
\begin{equation}
	M=R\Bigg[\dot{R}^2+1-\bigg(\frac{R}{L}\bigg)^2\Bigg]^{1/2}-R\Big(1-\frac{2m}{R}+\dot{R}^{2}\Big)^{1/2}.
	\label{11}
\end{equation}
The equation of motion of the shell is given by:
\begin{equation}
	\dot{M}+8\pi R\dot{R}p=-\dot{m}\dot{v}^3\;,
	\label{12}
\end{equation}
where $p$ is the pressure. Adopting the equation of state $p=(1-\gamma)\sigma$, then we can rewrite the equation of motion as,
\begin{equation}
	\dot{M}+8\pi R\dot{R}(1-\gamma)\sigma=-\dot{m}\dot{v}^3,
	\label{13}
\end{equation}
with $\sigma=M/4\pi R^2$.

The energy condition $\sigma+2p\ge 0$ with $p=(1-\gamma)\sigma$ is satisfied if $\gamma\le 3/2$ and energy condition $\sigma+p\ge0$ is satisfied if $\gamma\le 2$. Thus, the shell can be classified as normal matter for $\gamma\le 3/2$, as dark energy for  $3/2<\gamma\le 2$ and repulsive phantom energy shell for $\gamma>2$. The classification of matter for the thin shell, based on energy conditions, is summarized in Table 1:
\begin{table}[H]
	\centering
	\begin{tabular}{|c|c|c|c|}
		\hline
		\textbf{Matter} &\textbf{ EC 1} & \textbf{EC 2}&\textbf{ $\gamma$} \\
		\hline
		Normal Matter			  &  $\sigma+2p\ge 0$ &  $\sigma+p\ge 0$  & -1 or 0 \\
		Dark Energy			  &  $\sigma+2p< 0$   &  $\sigma+p\ge 0$ & 7/4 \\
		Repulsive Phantom Energy &  $\sigma+2p< 0$   &  $\sigma+p< 0$    & 3 \\
		Attractive Phantom Energy &  $\sigma+2p\ge 0$ &  $\sigma+p< 0$    &Impossible \\
		\hline
	\end{tabular}
	\caption{Matter classification of thin shell based on energy conditions (EC). The last column is the values of $\gamma$ that we considered in this section.}.
\end{table}

Since from the equation \eqref{13}, the thin shell mass, $M$ could depend on $\tau$ explicitly, then in general the equation \eqref{11} does not possess the Hamiltonian form as described in equation \eqref{1}. Therefore, in this paper, we assume
\begin{equation}
	\dot{m}\dot{v}^3=k_1(1-\gamma)\dot{M}+\gamma k_2\frac{\dot{R}}{R}M,
	\label{14}
\end{equation}
where $k_1$ and $k_2$ are arbitrary constants. We note that, for $\gamma = 0$, it revert back to the assumption made by R. Chan in \cite{chan2011radiating}. Substituting equation \eqref{14} into equation \eqref{13}, we obtain
\begin{equation}
	M=kR^{\frac{(2-k_2)\gamma-2}{k_1(1-\gamma)+1}}=kR^\eta,
	\label{15}
\end{equation} 
where $\eta\equiv\frac{(2-k_2)\gamma-2}{k_1(1-\gamma)+1}$ and $k$ is an integration constant. 

The interpretation of the first term in the equation \eqref{14} suggest that if $k_1 (1-\gamma)>0$, $\dot{M}$ will be positive, which denotes that $M$ decreases, and if $k_1 (1-\gamma)<0$, $\dot{M}$ will be negative, which means that $M$ increases due to mass transfer from the interior, consistent with results in \cite{chan2011radiating}. The second term implies that if $\gamma k_2>0$,  $\dot{R}$ will be negative, which indicates the decreasing of $R$ and if $\gamma k_2<0$, $\dot{R}$ will be positive, which represents the increasing of $R$. 

Substituting equation \eqref{15} to equation \eqref{11}, we obtain the Hamiltonian form of equation \eqref{11} with potential function:
\begin{equation}
	\begin{aligned}
	V=&-\frac{1}{8k^2L^4R^2R^{2\eta}}\big(-4R^{5}mL^2-4k^2R^2R^{2\eta}L^4+2k^2R^4R^{2\eta}L^2+4R^{2}m^2L^4\\
	&+4k^2RR^{2\eta}mL^4+R^{8}+k^4R^{4\eta}L^4\big). \label{16}
	\end{aligned}
\end{equation}
To eliminate integration constant $k$, we rescale the functions $m$, $L$ and $R$ as follows:
\begin{equation}\
	\begin{aligned}
		m &\to mk^{\frac{1}{1-\eta}},\\
		L &\to Lk^{\frac{1}{1-\eta}},\\
		R &\to Rk^{\frac{1}{1-\eta}}.
	\end{aligned}
	\label{17}
\end{equation}
Thus, the potential function becomes
\begin{equation}
	V=-\frac{1}{8L^4}\big(-4R^{3-2\eta}mL^2-4L^4+2R^2L^2+4R^{-2\eta}m^2L^4+4R^{-1}mL^4+R^{2(3-\eta)}+R^{2(\eta-1)}L^4\big). \label{18}
\end{equation}
The first derivative of the potential is given by,
\begin{equation}
	\begin{aligned}
		\frac{dV}{dR}=&-\frac{1}{8L^4}\big((8\eta-12)R^{2(1-\eta)}mL^2+4RL^2-8\eta R^{-2\eta-1}m^2L^4-4R^{-2}mL^4\\
		&+(6-2\eta)R^{5-2\eta}+(2\eta-2)R^{2\eta-3}L^4\big),
	\end{aligned}
	\label{19}
\end{equation}
and the solution to the $dV/dR = 0$ is
\begin{equation}
	\begin{aligned}
		m_{1,2}=&-\frac{1}{4\eta R^{-2\eta+1}L^2}\bigg(L^2-(2\eta-3)R^{2(2-\eta)}\\
		&\pm\sqrt{9R^{4(2-\eta)}+(6+4\eta)R^{2(2-\eta)}L^2+(4\eta^2-4\eta+1)L^4}\bigg).
	\end{aligned}
	\label{21}
\end{equation}
The second derivative of the potential is given by,
\begin{equation}
	\begin{aligned}
		\frac{d^2V}{dR^2}=&-\frac{1}{8L^4}\big((2-2\eta)(8\eta-12)R^{1-2\eta}mL^2+4L^2+8\eta(2\eta+1)R^{-2(\eta+1)}m^2L^4\\
		&+8R^{-3}mL^4+(5-2\eta)(6-2\eta)R^{2(2-\eta)}+(2\eta-3)(2\eta-2)R^{2(\eta-2)}L^4\big).
	\end{aligned}
	\label{20}
\end{equation}
In the next subsection, we consider the various cases for $k_1$ and $k_2$ and determine whether the possibility of formation of a stable gravastar. 

\subsection{\texorpdfstring{Case $k_1=k_2$}{Case k1=k2}}
In this subsection, we consider the cases where $k_1 = k_2$. Setting $k_1=k_2=+1$ and $k_1=k_2=-2$ yield a constant value for $\eta$ for all values of $\gamma$. For case $k_1=k_2=+1$, we get $\eta=-1$ across all values of $\gamma$, and similarly for $k_1=k_2=-2$ where $\eta=+2$ (see Table \ref{tableEtaIsotropic}). To generalize, we denote $\eta_{k_1,k_2}$ is the value of $\eta$ for given parameter $k_1$ and $k_2$.
\begin{table}[H]
	\centering
	\begin{tabular}{|c|c|c|}
		\hline
		$\mathbf{\gamma}$ & $\mathbf{\eta_{1,1}(k_1=k_2=+1)}$ & $\mathbf{\eta_{-2,-2}(k_1=k_2=-2)}$\\
		\hline
		 0   & -1 & +2 \\
		 -1  & -1 & +2 \\
		 7/4 & -1 & +2 \\
		 3   & -1 & +2 \\
		\hline
	\end{tabular}
	\caption{In case $k_1=k_2=+1$,  we always get $\eta=-1$ (denoted by $\eta_{1,1}$) for all values of $\gamma$ and in case $k_1=k_2=-2$, we always get $\eta=+2$ (denoted by $\eta_{-2,-2}$) for all values of $\gamma$.}
	\label{tableEtaIsotropic}
\end{table}

\subsubsection{\texorpdfstring{Case $k_1=k_2=+1$}{Case k1=k2=1}}
In this scenario, both the mass and the radius of the shell will decrease as a result of radiation. Given that $\eta = -1$ for all values of $\gamma$, and that the potential $V$ depends solely on $\eta$, it follows that the potential will remain consistent across different values of $\gamma$. By substituting $\eta = -1$ into equations \eqref{21}, \eqref{18}, and \eqref{20}, we can determine the expressions for mass $m$, potential $V$, and the second derivative of the potential $d^2V/dR^2$.

\begin{figure}[H]
	\centering
	\begin{subfigure}{0.48\textwidth}
		\includegraphics[width=\textwidth]{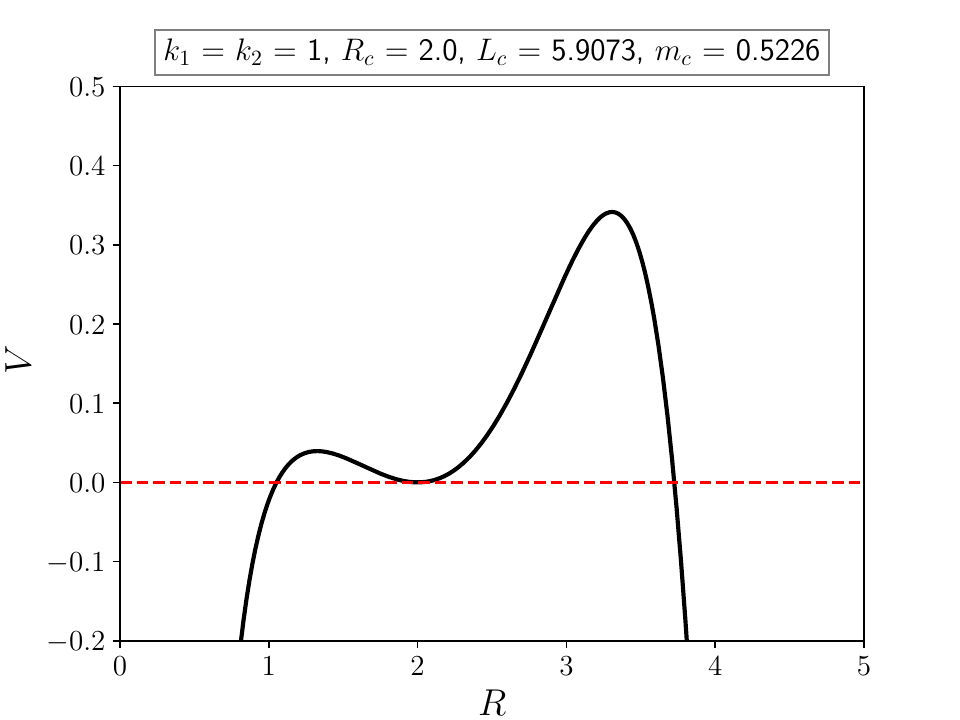}
		\caption{}
		\label{fig:potensialkp1a}
	\end{subfigure}
	\begin{subfigure}{0.48\textwidth}
		\includegraphics[width=\textwidth]{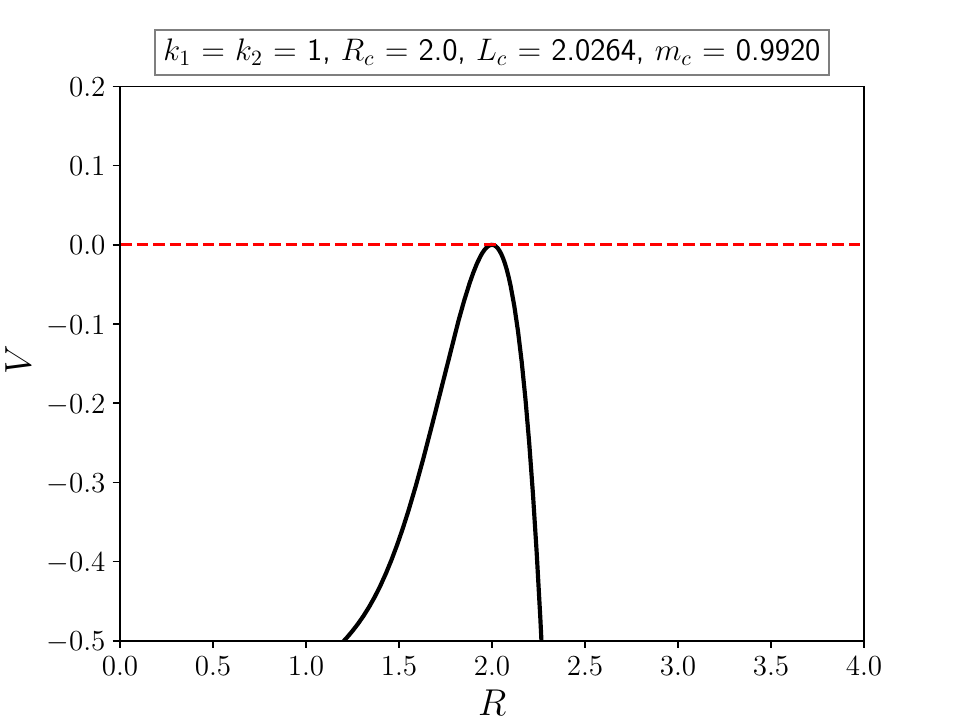}
		\caption{}
		\label{fig:potensialkp1b}
	\end{subfigure}
	\caption{The potential $V(R,m,L)$ for $k_1=k_2=+1$ and $R_c=2.0$. (a) For $m=m_1=m_c=0.5226$, $L_c=5,0973$, represent the formation of a stable gravastar. (b) For $m=m_2=m_c=0.9920$, $L_c=2.0264$, represent the formation of a black hole. }
\end{figure}

For the mass $m = m_1$, the second derivative of the potential can be either positive or negative, depending on the radius $R$ and the cosmological constant $L$. This variability suggests that the conditions necessary for both gravastar and black hole formation can be satisfied (see figure \ref{fig:potensialkp1a}). Conversely, for the mass $m = m_2$, the second derivative of the potential $d^2V/dR^2$ is always negative. This implies that in this case, the system can only lead to the formation of black holes (see figure \ref{fig:potensialkp1b}). 

\subsubsection{\texorpdfstring{Case $k_1=k_2=-2$}{Case k1=k2=-2}}
In this scenario, both the mass and the radius of the shell will increase since the source of radiation originates from the interior. Similar to the previous case, the potential will remain the same across different values of $\gamma$ since we only get $\eta = +2$ for all values of $\gamma$.

\begin{figure}[H]
	\centering
	\includegraphics[width=0.6\linewidth]{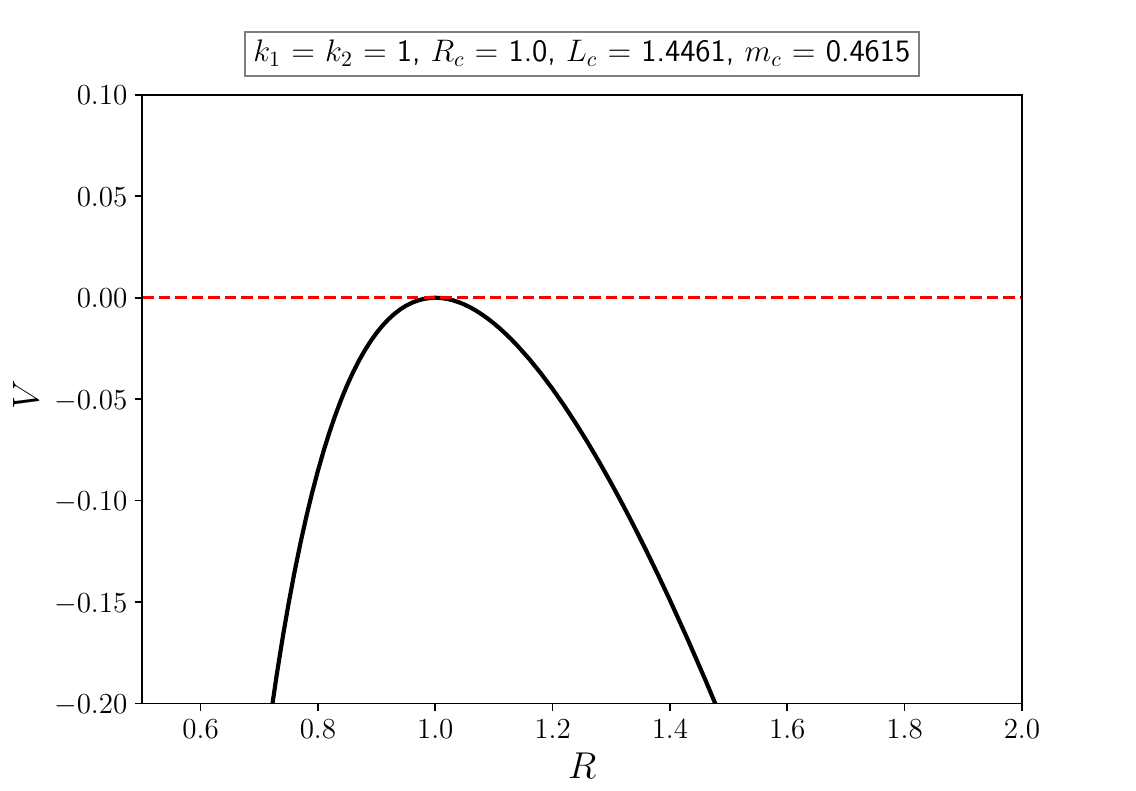}
	\caption{The potential $V(R,m,L)$ for $k_1=k_2=-2$ with $m=m_2=m_c=0.4615$, $R_c=1.0$ and $L_c=1.4461$. This represents the formation of a black hole.}
	\label{fig:potensialkn2}
\end{figure}

Given that the value of $m_1$ is always negative, the only viable solution is $m = m_2$. For the mass $m = m_2$, the second derivative of the potential is always negative, and will only lead to the formation of a black holes (see figure \ref{fig:potensialkn2}).

\subsection{\texorpdfstring{Case $k_1\ne k_2$}{Case k1 != k2}}

In this subsection, we consider the case where the values of $k_1$ and $k_2$ are different. In this case, we will observe different values of $\eta$ depending on $k_1$, $k_2$, and $\gamma$. This configuration allows us to examine the combined effects of changes in mass and radius on the potential. The values of $\eta$ for each case are summarized in Table \ref{Table:etaValues}.

\begin{table}[H]
	\centering
	\begin{tabular}{|c|c|c|}
		\hline
		 $\mathbf{\gamma}$ & $\mathbf{\eta_{1,-2}(k_1=+1, k_2=-2)}$ & $\mathbf{\eta_{-2,1}(k_1=-2,k_2=+1)}$ \\
		\hline
		 0   & -1  & +2     \\
		 -1  & -2  & +1     \\
		 7/4 & +20 & -1/10  \\
		 3   & -10 & +1/5   \\
		\hline
	\end{tabular}
	\caption{Values of $\eta$ for case $k_1=+1$, $k_2=-2$ and $k_1=-2$, $k_2=+1$. We get different value of $\eta$ for each case.}
	\label{Table:etaValues}
\end{table}

By substituting the value of $\eta$ into equations \eqref{21}, and \eqref{20}, we can determine the expressions for mass $m$ and the second derivative of the potential $d^2V/dR^2$. The possible signs for the masses dan the second derivative of the potential are summarized in Table \ref{Table:isotropicSummary}   
\begin{table}[H]
    \renewcommand{\arraystretch}{1.5}
	\centering
	\begin{tabular}{|c|c|c|c|c|c|}
		\hline
		$\mathbf{k_1}$ & $\mathbf{k_2}$ & $\mathbf{\gamma}$ &\multicolumn{2}{c|} {$\mathbf{m, \frac{d^2V}{dR^2} }$} & \textbf{Possible Formation}\\
		\hline
        \multirow{4}*{1} & \multirow{4}*{-2} & 0 & $m_1=+, \frac{d^2V_1}{dR^2}=\pm$ & $m_2=\pm, \frac{d^2V_2}{dR^2}=-$ & Gravastar/Black Hole \\ 
		\cline{3-6}
        ~ & ~ & -1 & $m_1=+,\frac{d^2V_1}{dR^2}=\pm$ & $m_2=\pm, \frac{d^2V_2}{dR^2}=-$ & Gravastar/Black Hole \\
		\cline{3-6}
        ~ & ~ & 7/4 & $m_1=\pm, \frac{d^2V_1}{dR^2}=\pm$ & $m_2=+, \frac{d^2V_2}{dR^2}=-$ & Gravastar/Black Hole \\
		\cline{3-6}
        ~ & ~ & 3 & $m_1=+, \frac{d^2V_1}{dR^2}=\pm$ & $m_2=\pm, \frac{d^2V_2}{dR^2}=-$ & Gravastar/Black Hole \\ 
		\hline
        \multirow{4}*{-2} & \multirow{4}*{1} & 0 & $m_1=-, \frac{d^2V_1}{dR^2}=-$ & $m_2=+, \frac{d^2V_2}{dR^2}=-$ & Black Hole \\
		\cline{3-6} 
        ~ & ~ & -1 & $m_1=-, \frac{d^2V_1}{dR^2}=-$ & $m_2=+, \frac{d^2V_2}{dR^2}=-$ & Black Hole \\
		\cline{3-6} 
        ~ & ~ & 7/4 & $m_1=+, \frac{d^2V_1}{dR^2}=\pm$ & $m_2=\pm, \frac{d^2V_2}{dR^2}=-$ & Gravastar/Black Hole \\
		\cline{3-6} 
        ~ & ~ & 3 & $m_1=-, \frac{d^2V_1}{dR^2}=\pm$ & $m_2=\pm, \frac{d^2V_2}{dR^2}=-$ & Black Hole \\
		\hline 
	\end{tabular}
	\caption{Summary of the masses and the second derivative of the potentials signs for case $k_1\ne k_2$.}
	\label{Table:isotropicSummary}
\end{table}

\subsubsection{Radiating Gravastars with Standard Matter Shell Type}
\begin{enumerate}
    \item Case $k_1 = +1$, $k_2 = -2$  ($\gamma = 0$ and $\gamma = -1$)\\
    In this scenario, we could consider the following cases:
	\begin{itemize}[label={--}]
		\item For $\gamma = 0$: The first term of equation \eqref{14} is positive ($k_1 (1-\gamma) > 0$) and the second term is zero ($\gamma k_2 = 0$). Therefore, the mass decreases without any change in the radius.
		\item For $\gamma = -1$: The first term remains positive ($k_1 (1-\gamma) > 0$) and the second term is also positive ($\gamma k_2 > 0$). Hence, both the mass and the radius decrease.
	\end{itemize}
	
    In this case, there are conditions under which we might find certain values of $R$ and $L$ with positive $m$. According to Table \ref{Table:isotropicSummary}, $m_1$ is always positive for both $\gamma = 0$ and $\gamma = -1$. This suggests the possibility of forming stable gravastars under these conditions.

    \item Case $k_1 = -2$, $k_2 = +1$  ($\gamma = 0$ and $\gamma = -1$)\\
    In this scenario, we could consider the following cases:
	\begin{itemize}[label={--}]
		\item For $\gamma = 0$: The first term of equation \eqref{14} is negative ($k_1 (1-\gamma) < 0$) and the second term is zero ($\gamma k_2 = 0$). Thus, the mass increases without any change in the radius.
		\item For $\gamma = -1$: The first term remains negative ($k_1 (1-\gamma) < 0$) and  the second term is also negative ($\gamma k_2 < 0$). Consequently, both the mass and the radius increase.
	\end{itemize}
   
    In this case, as shown in Table \ref{Table:isotropicSummary}, $m_1$ is negative and $m_2$ is positive for both $\gamma = 0$ and $\gamma = -1$. However, the second derivative of the potential is always negative for both masses, which lead to the formation of a black holes, and precluding the possibility of forming a stable gravastars.
\end{enumerate}

\begin{figure}[H]
	\centering
	\begin{subfigure}{0.48\textwidth}
		\includegraphics[width=\textwidth]{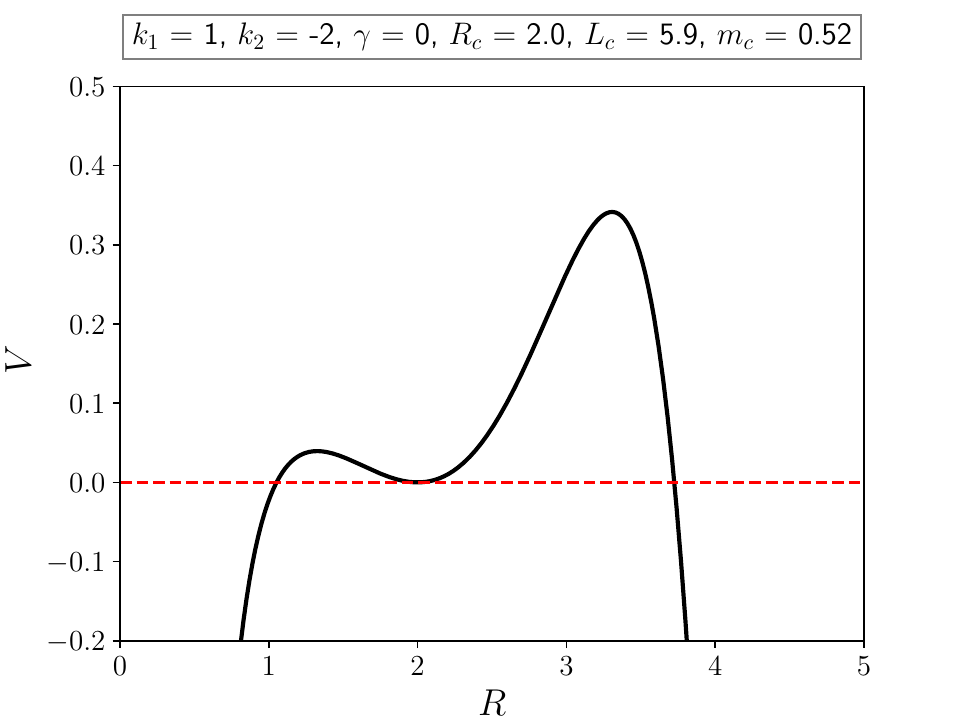}
		\caption{}
		\label{fig:potensialIsotropicStandard1a}
	\end{subfigure}
	\begin{subfigure}{0.48\textwidth}
		\includegraphics[width=\textwidth]{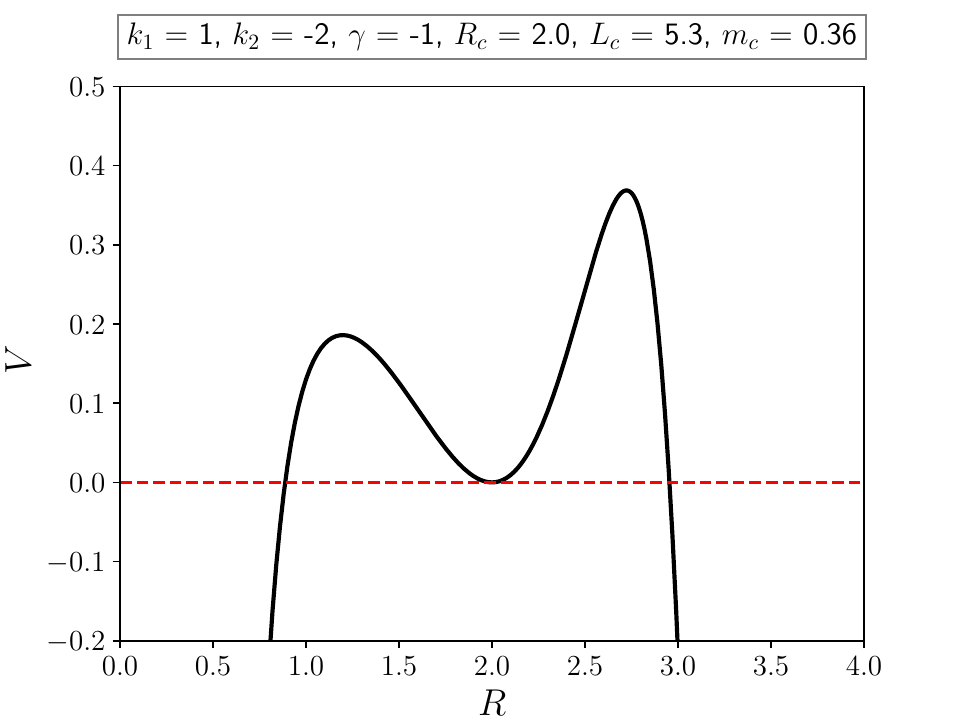}
		\caption{}
		\label{fig:potensialIsotropicStandard1b}
	\end{subfigure}
	\caption{The potential $V(R,m,L)$ for standar matter shell type with $k_1 = 1$, $k_2=-2$, and $R_c = 2$. (a) For $\gamma = 0$, $m_1 = m_c = 0.52$, and $L_c = 5.9$, and (b) for $\gamma = -1$, $m_1 = m_c = 0.36$, and $L_c = 5.3$. Both cases lead to a gravastar formation.}
\end{figure}

\subsubsection{Radiating Gravastars with Dark Energy Shell Type}
In exploring radiating gravastars with a dark energy shell, we examine two specific cases characterized by different parameter choices.

\begin{enumerate}
	\item Case $k_1=+1$, $k_2=-2$  and $\gamma=7/4$

	In this configuration, the first term in equation \eqref{14} is negative because $ k_1 (1 - \gamma) < 0 $. This indicates that the mass of the shell increase over time. The second term is also negative since \( \gamma k_2 < 0 \). Consequently, the radius of the shell is also increasing.

	From Table \ref{Table:isotropicSummary}, we observe that for $m_1$, the mass can be either positive or negative, and the second derivative of the potential can similarly be positive or negative. This suggests that for certain values of $ R $ and $ L $, the conditions for a stable gravastar can be satisfied. Therefore, it is possible to achieve stable gravastar formation at $ m = m_1 $ under this scenario. However, at $ m = m_2 $, although the mass is always positive, the second derivative of the potential is consistently negative. This outcome means that the shell configuration can only lead to the formation of black holes.

	\item Case $k_1=-2$, $k_2=+1$ and $\gamma=7/4$
	
	In this second case, the first term of equation \eqref{14} is positive ($k_1 (1-\gamma) > 0$) and the second term is positive ($\gamma k_2 > 0$). It means that mass and the radius both decreases.

	Table \ref{Table:isotropicSummary} shows that for $m_1$, the mass can again be either positive or negative, and the second derivative of the potential can be positive or negative. This means there are specific values of $R$ and $L$ where the conditions for a stable gravastar can be met. Consequently, stable gravastar formation is feasible at $m = m_1$ for this set of parameters. However, for $m = m_2$, despite the mass always being positive, the second derivative of the potential remains negative. As a result, this configuration is only leads to the formation of black holes.
\end{enumerate}

\begin{figure}[H]
	\centering
	\begin{subfigure}{0.48\textwidth}
		\includegraphics[width=\textwidth]{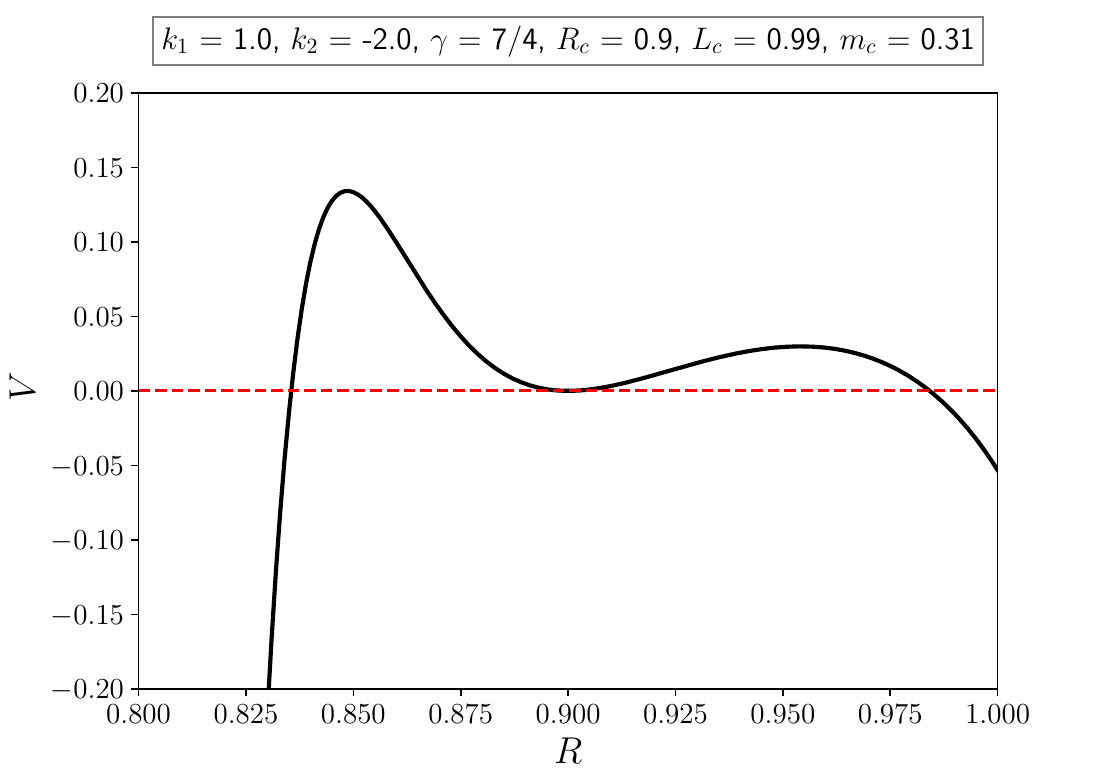}
		\caption{}
		\label{fig:potensialIsotropicDark1a}
	\end{subfigure}
	\begin{subfigure}{0.48\textwidth}
		\includegraphics[width=\textwidth]{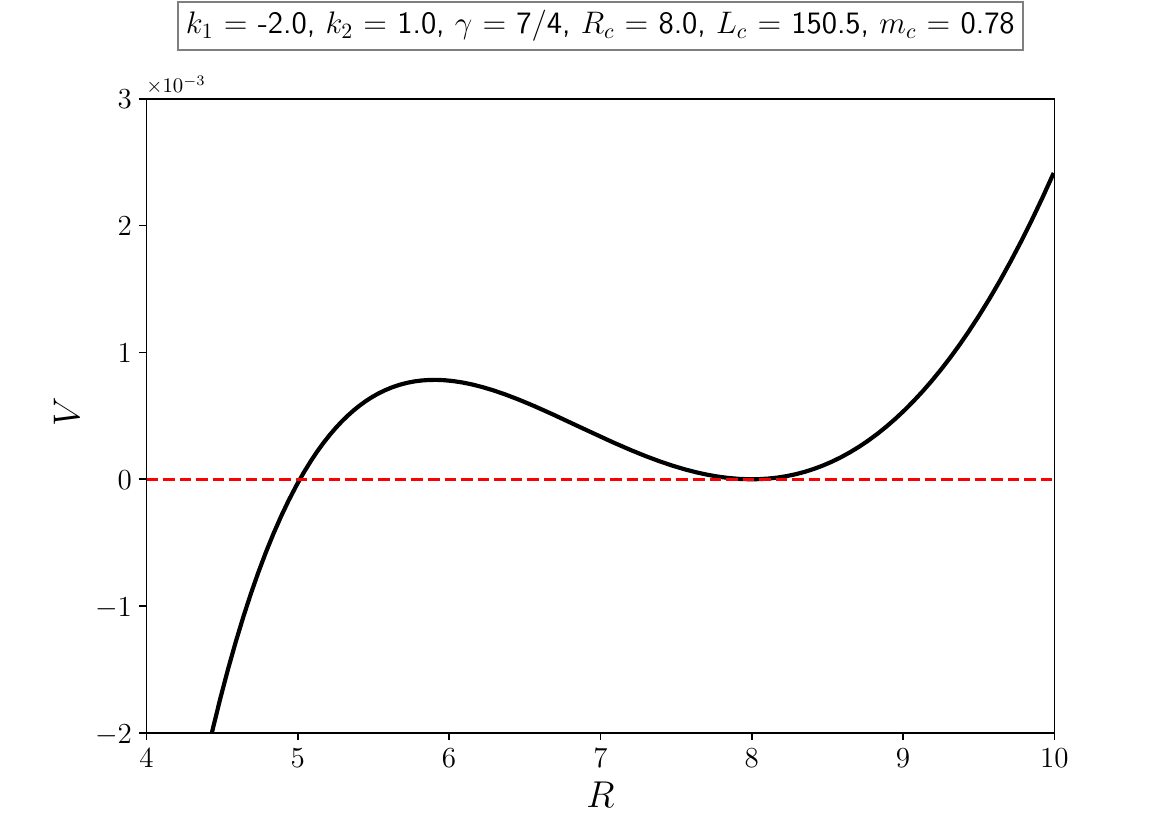}
		\caption{}
		\label{fig:potensialIsotropicDark1b}
	\end{subfigure}
	\caption{The potential $V(R,m,L)$ for dark energy shell type with $\gamma = 7/4$. (a) For $k_1 = 1$, $k_2=-2$, $R_c = 0.9$, $m_1 = m_c = 0.31$, and $L_c = 0.99$, and (b) for $k_1 = -2$, $k_2=1$, $R_c = 8.0$, $m_1 = m_c = 0.78$, and $L_c = 150.5$. Both cases lead to a gravastar formation.}
\end{figure}

\subsubsection{Radiating Gravastars with Repulsive Phantom Energy Shell Type}
\begin{enumerate}
	\item Case $k_1=+1$,  $k_2=-2$  and $\gamma=3$
	
	In this case, the first term in equation \eqref{14} is negative because $ k_1 (1 - \gamma) < 0 $. This indicates that the mass of the shell increase over time. The second term is also negative since \( \gamma k_2 < 0 \). Consequently, the radius of the shell is also increasing.
	
	Table \ref{Table:isotropicSummary} shows that $m_1$ is always positive while $m_2$ can be positive or negative. The second derivative of the potential at $m=m_1$ can be positive or negative and always negative at $m=m_2$. This implies that the gravastar formation is feasible at $m=m_1$. Meanwhile at $m=m_2$, it only lead to the formation of a black holes.
	
	\item Case $k_1=-2$,  $k_2=+1$ and $\gamma=3$
	
	In this second case, both first and second term of equation \eqref{14} are positive, $k_1 (1-\gamma)>0$ and $\gamma k_2>0$ respectively. It means that both the mass and the radius are decreasing, since in this case, the radiation comes directly from the shell.

	From the Table \ref{Table:isotropicSummary}, we observe that the mass $m_1$ is negative, thus the only possible solution is $m_2$ which can be positive or negative. However, the second derivative of the potential at $m_2$ is always negative. Hence, it only lead to the formation of a black holes.
\end{enumerate}

\begin{figure}[H]
	\centering
	\includegraphics[width=0.6\linewidth]{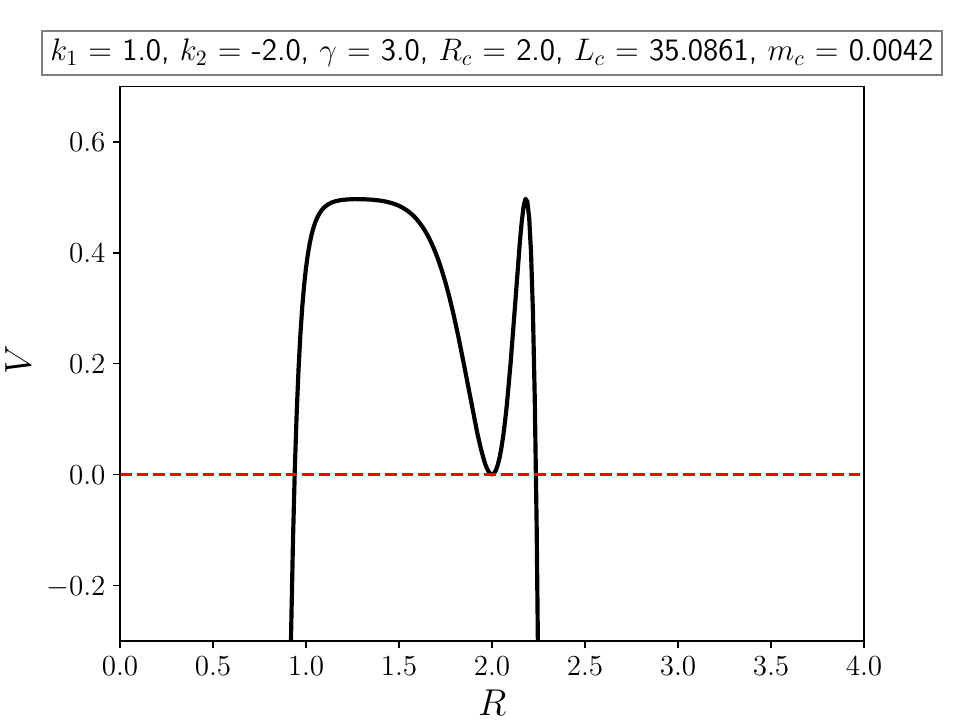}
	\caption{The potential $V(R,m,L)$ for repulsive phantom energy shell type with $\gamma = 3$. The formation of gravastar can be occurred for $k_1 = 1$, $k_2=-2$, $R_c = 2.0$, $m_1 = m_c = 0.0042$, and $L_c = 35.08$}
	\label{fig:potensialRepulsivePhantom}
\end{figure}


\subsection{Radiating Anti-de Sitter Gravastars}

Our mainly focus in this isotropic interior section is to study radiating gravastars with de Sitter interior which is a reguler gravastar model proposed. In this subsection, we just check and discuss in brief the possibility of existence of radiating anti-de Sitter (AdS) gravastars in the three-layer model. 

Following the metric form in the equation \eqref{4} but here $f_1=1+(r/L)^{2}$ and $f_{2}=[1+(r/L)^2]^{-1}=f_{1}^{-1}$. Then the mass of the thin shell will be

\begin{equation}
	M=R\Bigg[\dot{R}^{2}+1\Bigg(\frac{R}{L}\Bigg)^{2}\Bigg]^{1/2}-R\Big(1-\frac{2m}{R}+\dot{R}^{2}\Big)^{1/2}.
\end{equation}
By following the same steps as in de Sitter case (\eqref{12} to \eqref{17}), we obtain the potential of AdS interior case as
\begin{equation}
	V=-\frac{1}{8L^4}\big(4R^{3-2\eta}mL^2-4L^4-2R^2L^2+4R^{-2\eta}m^2L^4+4R^{-1}mL^4+R^{2(3-\eta)}+R^{2(\eta-1)}L^4\big). \label{Ads_18}
\end{equation}
The general form of the first derivative of the potential of AdS interior case can be written as
\begin{equation}
	\begin{aligned}
		\frac{dV}{dR}=&-\frac{1}{8L^4}\big((12-8\eta)R^{2(1-\eta)}mL^2-4RL^2-8\eta R^{-2\eta-1}m^2L^4-4R^{-2}mL^4\\
		&+(6-2\eta)R^{5-2\eta}+(2\eta-2)R^{2\eta-3}L^4\big).
	\end{aligned}
	\label{Ads_19}
\end{equation}
The general form of the second derivative of the potential of AdS interior case can be expressed as
\begin{equation}
	\begin{aligned}
		\frac{d^2V}{dR^2}=&-\frac{1}{8L^4}\big((2-2\eta)(12-8\eta)R^{1-2\eta}mL^2-4L^2+8\eta(2\eta+1)R^{-2(\eta+1)}m^2L^4\\
		&+8R^{-3}mL^4+(5-2\eta)(6-2\eta)R^{2(2-\eta)}+(2\eta-3)(2\eta-2)R^{2(\eta-2)}L^4\big).
	\end{aligned}
	\label{Ads_20}
\end{equation}
For $\frac{dV}{dR}=0$,
\begin{equation}
	\begin{aligned}
		m_{1,2}=&-\frac{1}{4\eta R^{-2\eta+1}L^2}\bigg(L^2-(3-2\eta)R^{2(2-\eta)}\\
		&\pm\sqrt{9R^{4(2-\eta)}-(6+4\eta)R^{2(2-\eta)}L^2+(4\eta^2-4\eta+1)L^4}\bigg).
	\end{aligned}
	\label{Ads_21}
\end{equation}
Here we do not find stable radiating anti-de Sitter gravastar formation with the combination of $k_{1}$ and $k_{2}$ as we apply in the de Sitter interior case. Hence, we try to explore several values of $k_{1}$ and $k_{2}$ which can be possibly produce gravastar. We can find a few of them which are provided in Table \ref{TableEtaIsotropicAdS}.

\begin{table}[H]
	\centering
	\begin{tabular}{|c|c|c|c|}
		\hline
		$\mathbf{\gamma}$ & $\mathbf{k_{1}}$ & $\mathbf{k_{2}}$ & $\mathbf{\eta}$\\
		\hline
		0   & -2.3 & 1 & 1.5385\\
		-1  & -0.8 & 3 & 1.6667\\
		7/4 & 0.8 & 0.5 & 1.5625 \\
		3   & -2.7 & -2.5 & 1.7969 \\
		\hline
	\end{tabular}
	\caption{Several values of $k_{1}$, $k_{2}$ and $\eta$ that produce radiating anti-de Sitter gravastar formations for various shell types.}
	\label{TableEtaIsotropicAdS}
\end{table}

\subsubsection {Radiating AdS Gravastars with Standard Energy Shell Type}
	
	As in de Sitter case, we take $\gamma=0$ and $\gamma=-1$ for standard thin shell. We can see that from \eqref{15}, for $\gamma=0$, $\eta$ will only depend on $k_{1}$. Here, we take $k_{1}=-2.3$ and just take $k_{2}=1$ as values which can produce stable AdS gravastars. Meanwhile, for $\gamma=-1$, we obtain that the value of $k_{1}=-0.8$ and $k_{2}=3$ can give stable radiating AdS gravastars.
	
	In this case, $k_1(1-\gamma)<0$ for both $\gamma=0$ and $\gamma=-1$ which is interpreted that the mass of the gravastar increases. Meanwhile for $\gamma=0$ clearly radius does not change and for $\gamma=-1$, radius increases since $\gamma k_{2}<0$.
	
	\begin{figure}[H]
		\centering
		\begin{subfigure}{0.48\textwidth}
			\includegraphics[width=\textwidth]{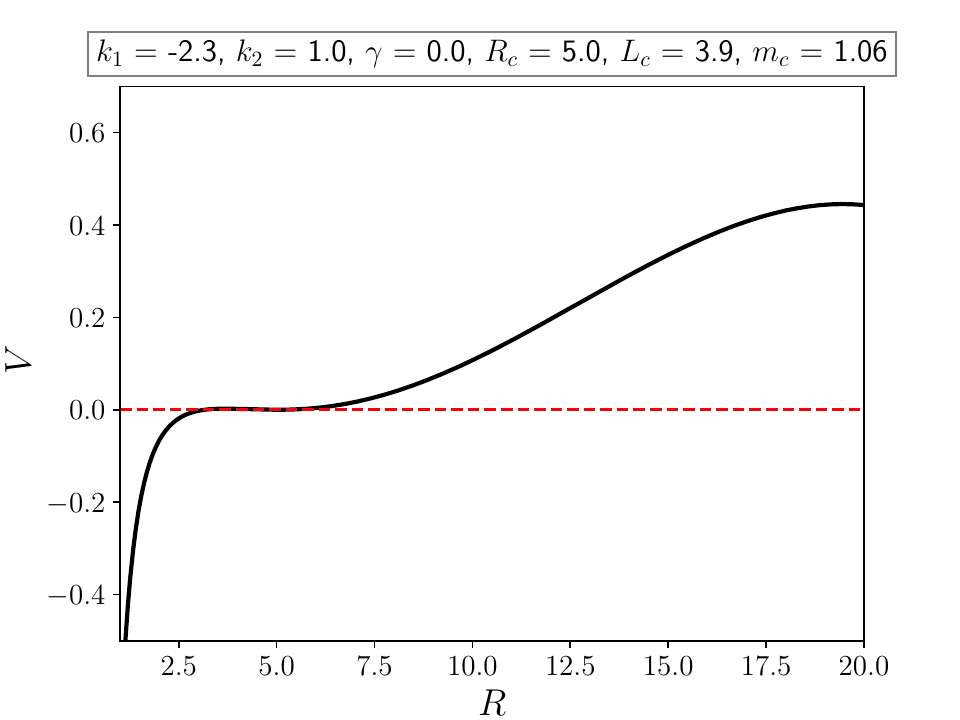}
			\caption{}
			\label{fig:potensialadsg0}
		\end{subfigure}
		\begin{subfigure}{0.48\textwidth}
			\includegraphics[width=\textwidth]{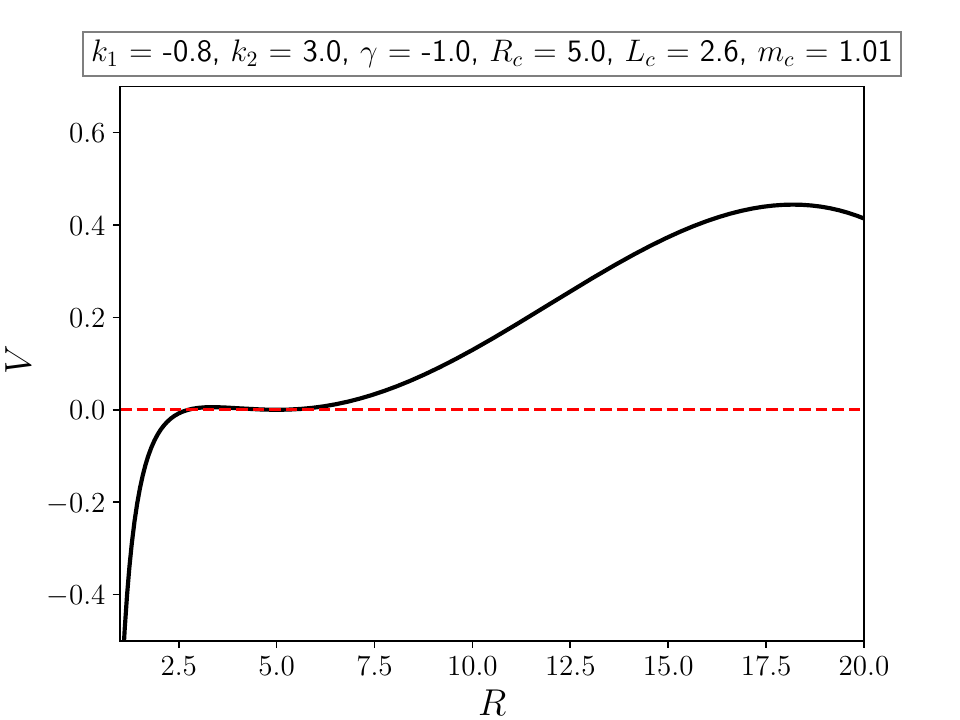}
			\caption{}
			\label{fig:potensialgadsn1}
		\end{subfigure}
		\caption{The potential $V(R,m,L)$ of AdS interior case for standard energy shell type with (a) $\gamma=0$ and (b) $\gamma=-1$. For (a), the formation of radiating AdS gravastar can be occurred when $k_{1}=-2.3$, $k_{2}=1$, $R_c=5.0$, $L_{c}=3.924285$, and $m_{c}=1.058613$. For (b), the formation of radiating AdS gravastar can be occurred when $k_{1}=-0.8$, $k_{2}=3$, $R_c=5.0$, $L_{c}=2.623686$, and $m_{c}=1.010552$. }
	\end{figure}

\subsubsection{Radiating AdS Gravastars with Dark Energy Shell Type}
	
	We take $\gamma=7/4$ for dark energy thin shell. We can obtain stable gravastars in this case for $k_{1}=0.8$ and $k_{2}=0.5$. Consequently, in this case, $k_1(1-\gamma)<0$ which means that the mass increases and $\gamma k_{2}>0$ which means that the radius decreases.
	
\begin{figure}[H]
	\centering
	\includegraphics[width=0.6\linewidth]{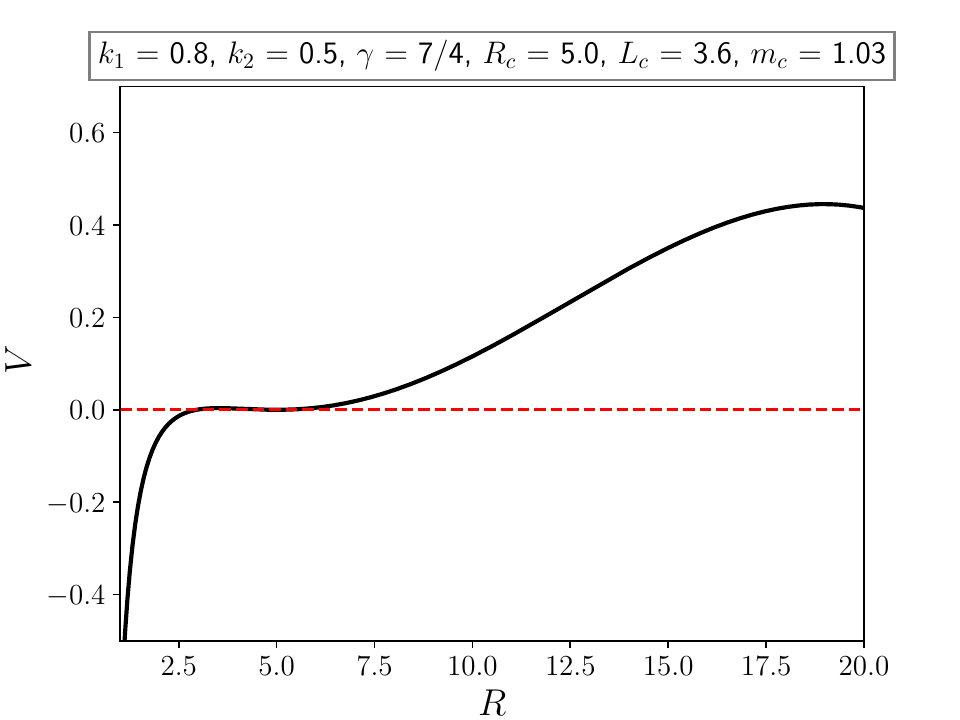}
	\caption{The potential $V(R,m,L)$ of AdS interior case for dark energy shell type. For $\gamma=7/4$, the formation of radiating AdS gravastar can be occurred when $k_{1}=0.8$, $k_{2}=0.5$, $R_c=5.0$, $L_{c}=3.614758$, and $m_{c}=1.033653$.}
	\label{fig:lgg74k108k205}
\end{figure}
	
	\subsubsection {Radiating AdS  Gravastars with Repulsive Phantom Energy Shell Type}
	
	We take $\gamma=3$ for repulsive phantom energy thin shell. We can find AdS stable gravastars in this case for $k_{1}=-2.7$ and $k_{2}=-2.5$. By taking these values, we obtain that $k_1(1-\gamma)>0$ which means that the mass decreases and $\gamma k_{2}<0$ which means that the radius increases.
	
\begin{figure}[H]
	\centering
	\includegraphics[width=0.6\linewidth]{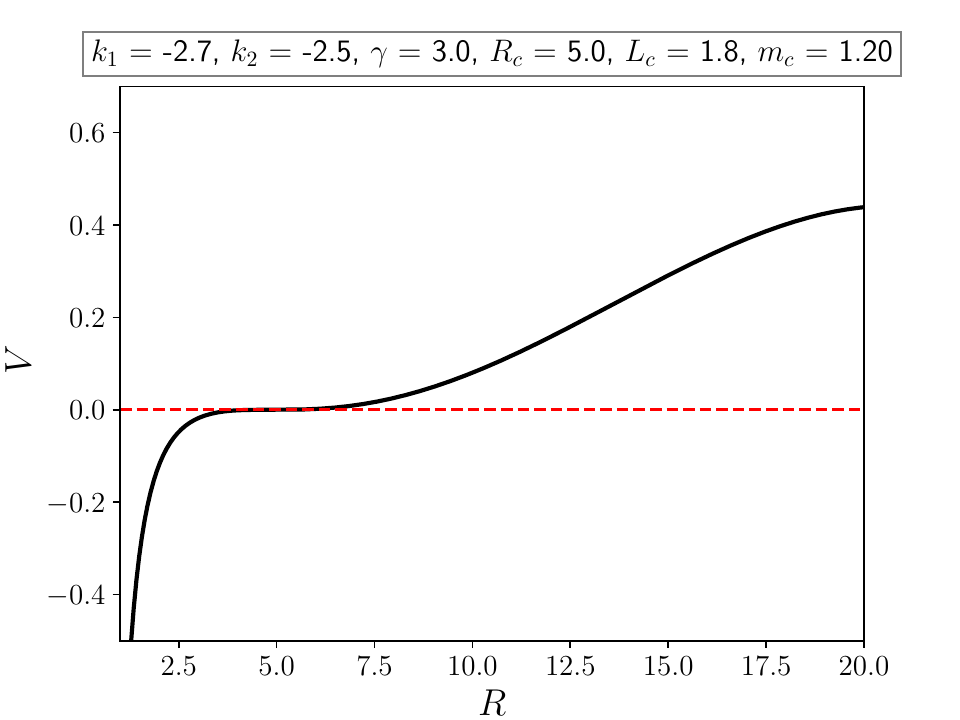}
	\caption{The potential $V(R,m,L)$ of AdS interior case for repulsive phantom shell type. For $\gamma=3$, the formation of radiating AdS gravastar can be occurred when $k_{1}=-2.7$, $k_{2}=-2.5$, $R_c=5.0$, $L_{c}=1.847214$, and $m_{c}=1.203552$.}
	\label{fig:lgk1n27k2n25g3}
\end{figure}

We note here that one of differences between the de Sitter gravastars and the anti-de Sitter ones is comparison between value of $R_{c}$ and $L_c$. The de Sitter gravastars have $R_c < L_c$ while anti de sitter ones have $R_c > L_c$. This is related to the metric. For de Sitter gravastars, it must be $2m < R < L$ to avoid any horizon formed whereas for anti de Sitter interior, it can be $R > L$.


\section{Anisotropic Interior}
\label{sec:anisotropic}

\subsection{Gravastar Model for Anisotropic Interior Case}
In this section, we consider a model of gravastar with anisotropic interior which can be written generally in the form,
\begin{equation}
	ds^{2}_i=-\exp\bigg[-2\int_{r}^{\infty}g(\tilde{r})d\tilde{r}\bigg]dt^{2}+\frac{dr^{2}}{1-\frac{2\bar{m}(r)}{r}}+r^{2}(d\theta^{2}+\sin^{2}\theta d\phi^{2}),
\end{equation}
where $g(r)$ is called "gravity profile". Positive $g(r)$ means inwardly gravitational attraction and negative $g(r)$ means outward gravitational repulsion.

In this paper, we consider a Tolman-Matese-Whitman (TMW) model which has the gravity profile function as \cite{lobo2006stable},
\begin{equation}
	g(r)=\bigg(\frac{br}{2}\bigg)\bigg[\frac{(1+3\omega)+(1+\omega)2br^2}{(1+br^2)(1+2br^2)}\bigg],
\end{equation} 
and the mass function,
\begin{equation}
	\bar{m}(r)=\frac{br^3}{2(1+2br^2)},
\end{equation}
where $b$ is an arbitrary positive constant and $\omega$ is a constant that defines the equation of state of the gravastar interior. We note that $g(r)<0$ can be obtained if $\omega<\omega_{c}$ with $\omega_{c}\equiv \sup_{r_0\in(0,\infty)} \;-(1+2br_{0}^{2})/(3+2br_{0}^{2})$. The energy density $\rho$, the radial pressure $p_r$ and the tangential pressure $p_t$ for the interior are given by
\begin{equation}
	\begin{aligned}
		p_r&=\omega \rho=\bigg(\frac{\omega b}{8\pi}\bigg)\bigg(\frac{3+2br^2}{(1+2br^2)^2}\bigg),\\
		p_t&=\bigg(\frac{\omega b}{8\pi}\Big)\bigg(\frac{3+2br^2}{(1+2br^2)^2}\bigg)+\frac{b^2 r^2}{32\pi[(1+2br^2)^3(1+br^2)]}\times \\
		&\Big((1+\omega)(3+2br^2)[(1+3\omega)+2br^2(1+\omega)]-8\omega(5+2br^2)(1+br^2)\Big).
	\end{aligned}
\end{equation}

We would like to note that we have corrected the formulation of $p_t$ by changing the negative sign in the first term (as used in the previous work of R. Chan et al. \cite{chan2009stable}) to a positive sign. This correction results in a classification that differs from the findings of R. Chan et al. \cite{chan2009stable}. Our corrected results align with the classification method proposed by Lobo \cite{lobo2006stable}, as well as with the classification method of R. Chan et al., which is based on energy conditions (as explained in subsection 3.2).

Thus, we can write the interior metric of the gravastar as,
\begin{equation}
	ds_{i}^2=-\tilde{f_1} dt^2+\tilde{f_2} dr^2+r^2 (d\theta^2+\sin^2\theta d\phi^2),
\end{equation}
where
\begin{equation}
	\begin{aligned}
		\tilde{f_1}&=(1+br^2)^{\frac{1-\omega}{2}}(1+2br^2)^{\omega},\\
		\tilde{f_2}&=\frac{1+2br^2}{1+br^2},
	\end{aligned}
\end{equation}
Meanwhile the exterior region is Vaidya spacetime as metric given by equation \eqref{5},
\begin{equation*}
	ds_e^2=-Fdv^2-2d\bar{r}dv+\bar{r}^2(d\theta^2+\sin^2\theta d\phi^2),
\end{equation*}
where $F=1 - 2m(v)/\bar{r}$.  

The mass of the shell can be written as,
\begin{equation}
	\label{30}
	M=R\bigg(\frac{1+bR^2+(1+2bR^2)\dot{R}^2}{(1+bR^2)^{(-1+\omega)/2}(1+2bR^2)^{\omega+2}}\bigg)^{1/2}-R\Big(1-\frac{2m}{R}+\dot{R}^{2}\Big)^{1/2}.
\end{equation}
The thin shell equation of motion of radiating gravastars generally is given by \cite{chan2011radiating}
\begin{equation}
	\label{32}
	\dot{M}+8\pi R\dot{R}(1-\gamma)\sigma=-\dot{m}\dot{v}^{3},
\end{equation}
where $\sigma=M/4\pi R^{2}$ and we adopting the equation of state $p=(1-\gamma)\sigma$ in the thin shell. 

Similar with the isotropic case, the equation \eqref{30} does not possess the Hamiltonian form as described in equation \eqref{1}. Therefore, in the rest of this paper, we assume
\begin{equation}
	\label{33}
	\dot{m}\dot{v}^{3}=k_1(1-\gamma)\dot{M},
\end{equation}
where $k_1$ is an arbitrary constant. We note that, for $\gamma = 0$, it revert back to the assumption made by R. Chan in \cite{chan2011radiating}. Substituting assumption \eqref{33} to equation \eqref{32}, we will get thin shell mass $M$,
\begin{equation}
	\label{34}
	M=kR^{\frac{2(\gamma-1)}{(1-\gamma)k_1+1}}\equiv kR^{\eta},
\end{equation}
where $\eta\equiv 2(\gamma-1)/[(1-\gamma)k_1+1]$ and $k$ is an integration constant. Substituting equation \eqref{34} to equation \eqref{30}, and rescaling $m$, $L$ and $R$ as,
\begin{equation}
	\begin{aligned}
		m\to mk^{\frac{1}{1-\eta}},\\
		b\to bk^{-\frac{2}{1-\eta}},\\
		R\to Rk^{\frac{1}{1-\eta}},
	\end{aligned}
\end{equation}
we get the potential in the form of equation \eqref{1} which can be written as,
\begin{equation}
	\begin{aligned}
		V(R,m,b,\omega, \eta)=&-\frac{1}{2R^{2}b_2\big(b_2^{\omega+1}-b_1^{\frac{\omega+1}{2}}\big)^{2}}\Big(b_2^{\omega+2}b_1^{\frac{\omega+1}{2}}R^{2\eta}\\
		&-2R^{\eta}b_2^{\frac{3\omega+4}{2}}b_1^{\frac{\omega+1}{4}}\Big[b_2^{-\omega}b_1^{\frac{\omega+1}{2}}R^2-b_2^{-(\omega+1)}b_1^{\frac{\omega+3}{2}}R^2-2b_2^{-\omega}b_1^{\frac{\omega+1}{2}}mR+b_1R^2\\
		&-b_2R^2+2b_2mR+b_2R^{2\eta}\Big]^{\frac{1}{2}}+b_2^{\omega+2}b_1^{\frac{\omega+1}{2}}R^2-b_2^{2\omega+3}R^2-2b_2^{\omega+2}b_1^\frac{\omega+1}{2}mR\\
		&+2b_2^{2\omega+3}mR+b_2^{2\omega+3}R^{2\eta}-b_1^{\omega+2}R^2+b_2^{\omega+1}b_1^{\frac{\omega+3}{2}}R^2\Big),
	\end{aligned}
\end{equation}
where
\begin{equation}
	b_1\equiv 1+bR^2, \quad b_2\equiv 1+2bR^2.
\end{equation}

In their study, R. Chan et al. offer a classification of matter in both the interior and the thin shell of gravastars based on energy conditions \cite{chan2009stable}. They categorize dark energy as a type of fluid that violates the strong energy condition (SEC). Additionally, they define phantom energy as a fluid that violates at least one of the null energy conditions (NECs). The phantom energy is further subdivided into two types, namely Repulsive Phantom Energy, in which violates the SEC, leading to repulsive gravitational effects, and Attractive Phantom Energy, in which satisfies the SEC, thus resulting in attractive gravitational effects. The classification of interior matter is summarized in Table \ref{Table:AnisotropicInteriorMatter},
\begin{table}[H]
	\centering
	\begin{tabular}{|c|c|c|c|}
		\hline
		\textbf{Matter} & \textbf{EC 1} & \textbf{EC 2}& \textbf{EC 3} \\
		\hline
		Normal Matter			  &  $\rho+p_r+2p_t\ge 0$ &  $\rho+p_r\ge 0$  & $\rho+p_t\ge 0$  \\
		Dark Energy 			  &  $\rho+p_r+2p_t< 0$ &  $\rho+p_r\ge 0$  & $\rho+p_t\ge 0$  \\
		Repulsive Phantom Energy &  $\rho+p_r+2p_t< 0$ &  $\rho+p_r< 0$  & $\rho+p_t\ge 0$  \\
		Repulsive Phantom Energy  &$\rho+p_r+2p_t< 0$ &  $\rho+p_r\ge 0$  & $\rho+p_t< 0$  \\
		Repulsive Phantom Energy  &  $\rho+p_r+2p_t< 0$ &  $\rho+p_r< 0$  & $\rho+p_t< 0$  \\
		Attractive Phantom Energy &  $\rho+p_r+2p_t\ge 0$ &  $\rho+p_r< 0$  & $\rho+p_t\ge 0$  \\
		Attractive Phantom Energy  &  $\rho+p_r+2p_t\ge 0$ &  $\rho+p_r\ge 0$  & $\rho+p_t< 0$  \\
		Attractive Phantom Energy  &  $\rho+p_r+2p_t\ge 0$ &  $\rho+p_r< 0$  & $\rho+p_t< 0$  \\
		\hline
	\end{tabular}
	\caption{Interior matter classification summary based on energy conditions (EC) where we assume that $\rho\ge 0$ \cite{chan2009stable}.}
	\label{Table:AnisotropicInteriorMatter}
\end{table}
We also consider several cases of equation of state in the interior which described by Table \ref{Table:EqOfStateSummary}. In the subsequent sections, we will apply these classification schemes to analyze and interpret the possible formation of a stable gravastars.
\begin{table}[H]
	\centering
	\begin{tabular}{|c|c|c|}
		\hline
		\textbf {Case} & $\mathbf{\omega}$ & \textbf{b} \\
		\hline
		A  & 0.1           & 0.0001\\
		B  & 1.5           & 0.01\\
		C  & -0.5          & 0.01\\
		D  & -0.7          & 0.01\\
		E  & -1.25         & 0.01\\
		F  & -1.5          & 0.01\\
		\hline
	\end{tabular}
	\caption{We take some interior parameters $\omega$ and $b$ to be analyzed. The cases A \& B, C \& D and E \& F are standard, dark and repulsive phantom energy, respectively.}
	\label{Table:EqOfStateSummary}
\end{table}


\subsection{Standard Interior Case}
A stable radiating star with a standard interior is found only in the case of $k_1 = +1$, with both standard and repulsive phantom thin shells. For $k_1 = -2$, a standard thin shell always results in a black hole, while with non-standard (dark and repulsive phantom) thin shells, no stable structure is found. Here, we consider two combinations of $\omega$ and $b$ as examples, providing only the most significant graphs for efficiency. The first combination is $\omega = 1.5$ and $b = 0.01$, and the second is $\omega = 0.1$ and $b = 0.0001$.

\begin{figure}[H]
	\centering
	\includegraphics[width=0.6\linewidth]{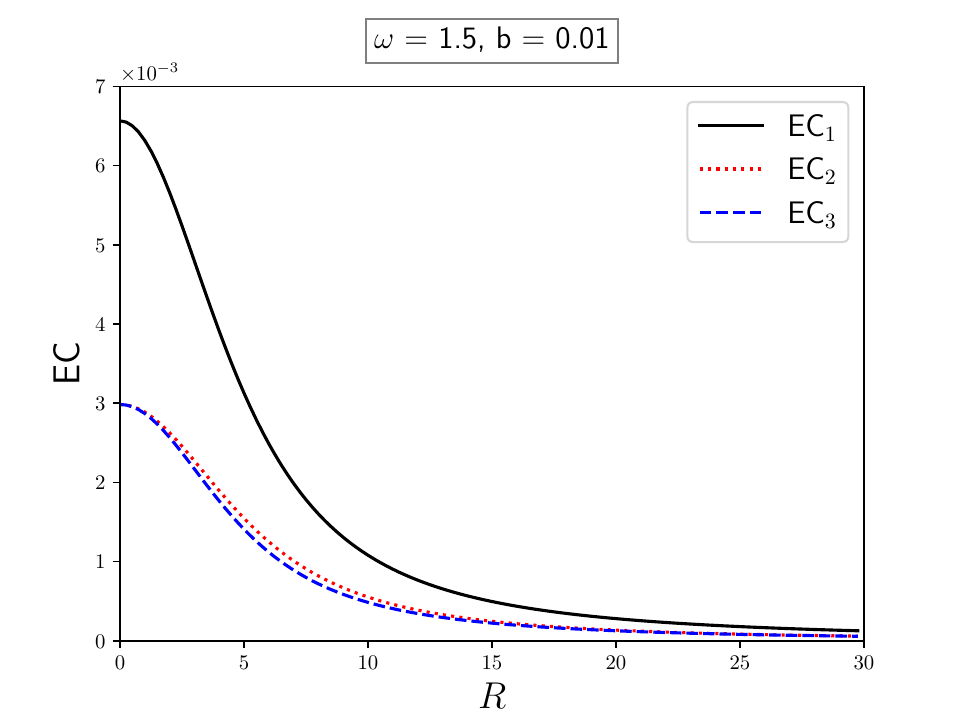}
	\caption{Energy condition $\text{EC}_1\equiv \rho+p_r+2p_t$, $\text{EC}_2\equiv \rho+p_r$ and $\text{EC}_3\equiv \rho+p_t$ for $\gamma=1.5$ and $b=0.01$. We can see that $\text{EC}_1$, $\text{EC}_2$ and $\text{EC}_3$ are positive for all values of $R$ which means that this is standard energy interior case.}
	\label{fig:fp}
\end{figure}

We demonstrate that an interior with $\omega = 1.5$ and $b = 0.01$ is classified as a standard interior (see Figure \ref{fig:fp}), contrary to the classification as a repulsive phantom interior shown by R. Chan et al. in \cite{chan2009stable}. This discrepancy arises due to the sign difference in the first term of $p_t$, as noted in subsection 3.1.

\begin{table}[H]
	\begin{tabular}{|c|c|c|c|c|c|c|c|}
		\hline
		\textbf{Case}&$\mathbf{\gamma}$&\multicolumn{2}{c|}{$\mathbf{k_1=0}$} & \multicolumn{2}{c|}{$\mathbf{k_1=+1}$} & \multicolumn{2}{c|}{$\mathbf{k_1=-2}$} \\
		\hline
		\multicolumn{2}{|c|}{}&$\mathbf{m_c}$ & $\mathbf{R_c}$ & $\mathbf{m_c}$ & $\mathbf{R_c}$ & $\mathbf{m_c}$  & $\mathbf{R_c}$ \\
		\hline
		A &-1 &0.012004  & 3.732847  & 0.117444 &6.515841  & 1.305340 & 4.066782 \\
		
		& 0 & 0.054698 & 5.505893 & 0.185319 & 7.078525 & 0.592488  & 1.333059 \\
		
		& 7/4 &0.842998 & 2.246995 &0.504560 & 1.017530 & -  & - \\
		
		& 3 & 0.512068 & 1.045473 &0.012004  & 3.732847 &- &- \\
		\hline
		B &-1 & 0.209764 & 1.799113 & 0.487724 & 1.919223  & 0.958161 & 2.628539\\
		
		&0  &0.378028  & 1.959559  & 0.553889 & 1.764609 & 0.575614 & 1.289608 \\
		
		& 7/4 & 0.751538 & 1.919813 & 0.502354 & 1.013288 & -  & -\\
		
		& 3 &0.508339 & 1.038145 &  0.209764 & 1.799113 & - & -   \\
		\hline	
	\end{tabular}
	\caption{Comparison of critical mass $m_c$ and critical radius $R_c$ between the radiating case ($k_1=+1$, $k_1=-2$) and the non-radiating case ($k_1=0$) with standard energy interior. Case A: $\omega=0.1$ and $b=0.0001$ and case B: $\omega=1.5$ and $b=0.01$.}
	\label{Table:StandardInterior}
\end{table}

\begin{figure}[H]
	\centering
	\begin{subfigure}{0.48\textwidth}
		\includegraphics[width=\textwidth]{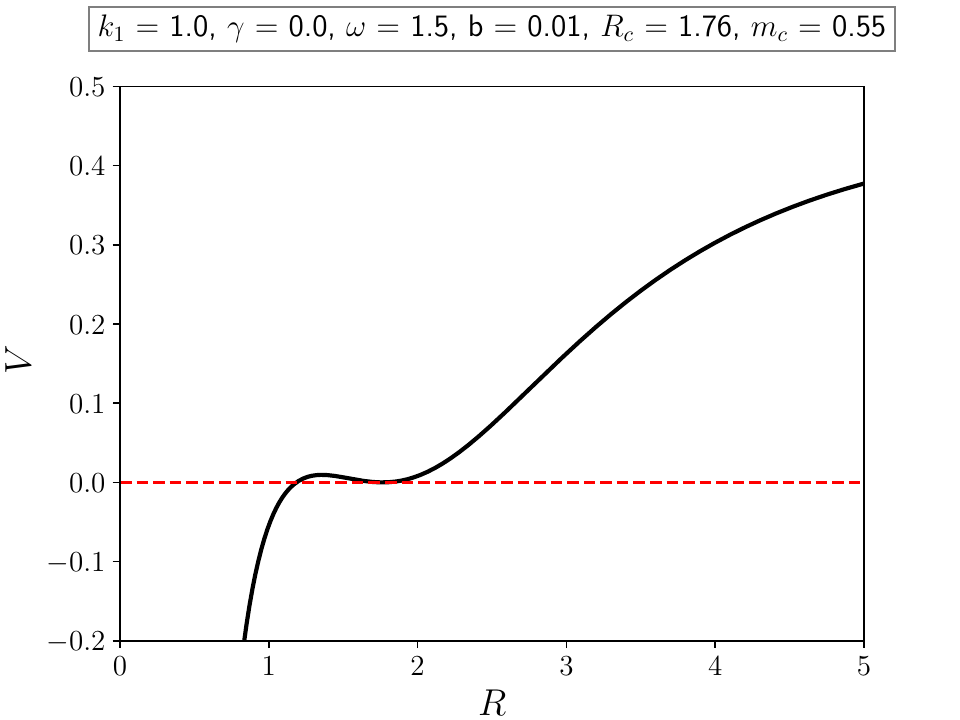}
		\caption{}
	\end{subfigure}
	\begin{subfigure}{0.48\textwidth}
		\includegraphics[width=\textwidth]{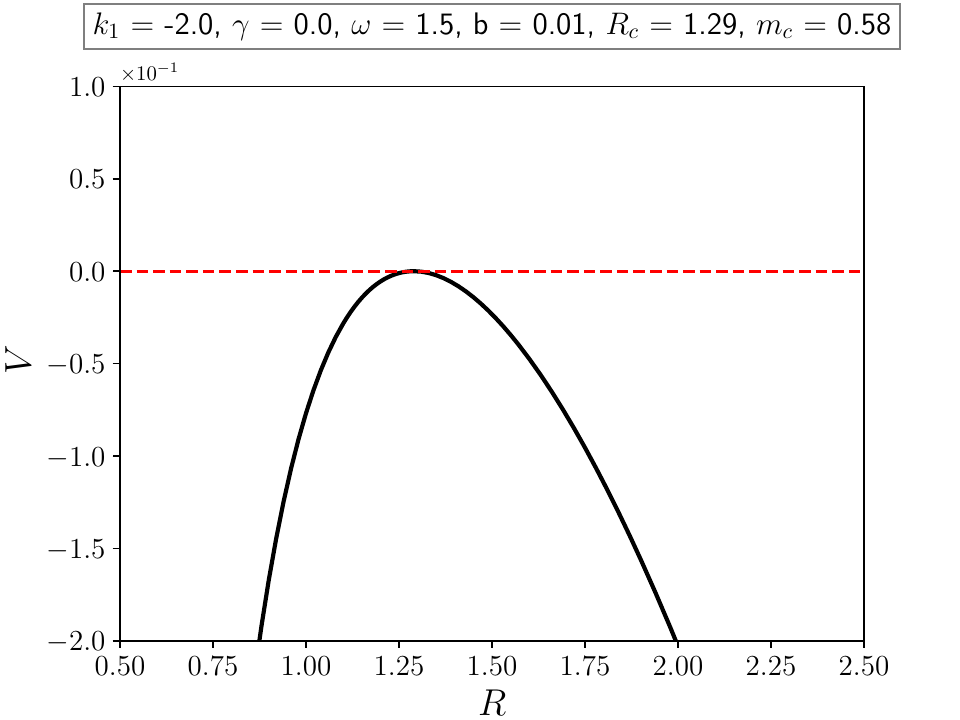}
		\caption{}
	\end{subfigure}
	\caption{The potential $V(R)$ for $\gamma=0$, $\omega=1.5$, and $b=0.01$ with (a) $k_1=+1$ and (b) $k_1=-2$. We note that for $k_1=+1$ case, stable gravastar can be found, while for $k_1=-2$, only black hole is found.}
\end{figure}

\subsection{Dark Interior Case}

Following the classification performed by Lobo \cite{lobo2006stable} and R. Chan et al. \cite{chan2009stable}, dark energy can be identified in the range $-1/3 > \omega > -1$. We consider two combinations of $\omega$ and $b$, where $\omega$ remains within this range. Notably, this type of interior could not be identified by R. Chan et al. due to the use of a negative sign in the first term of $p_t$.

\begin{figure}[H]
	\centering
	\includegraphics[width=0.6\linewidth]{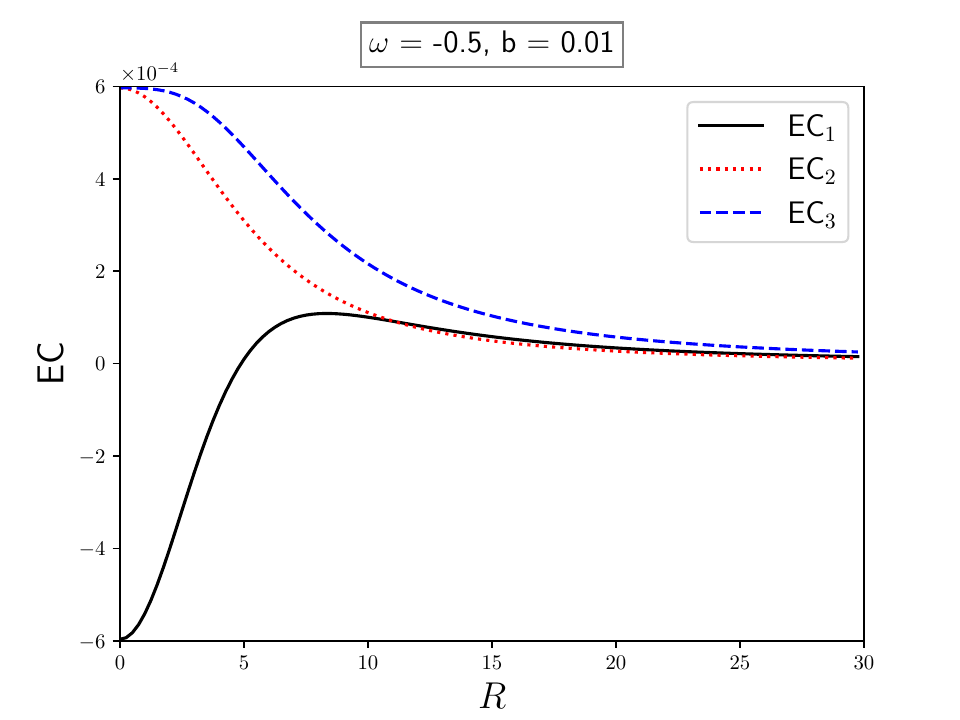}
	\caption{Energy condition $\text{EC}_1\equiv \rho+p_r+2p_t$, $\text{EC}_2\equiv \rho+p_r$ and $\text{EC}_3\equiv \rho+p_t$ for $\gamma=-0.5$ and $b=0.01$. We can see that for small $R$ ($R\leq R_c$) $\text{EC}_1$ is negative, while $\text{EC}_2$ and $\text{EC}_3$ are positive which means that this is dark energy interior case.}
	\label{fig:dp}
\end{figure}

In this analysis, we take $\omega = -0.5$ and $\omega = -0.7$, with $b = 0.01$ for both values of $\omega$. The values of $R_c$ and $m_c$ for these cases are listed in Table \ref{Table:DarkInterior}. Since $R_c$ represents the radius where the collapsing star is static and stable, it can be interpreted as the radius of the gravastar. We verify that this gravastar radius satisfies the condition $g(r) < 0$ ($\omega < \omega_c$), indicating a repulsive interior. Notably, $\omega_c$ is larger for smaller values of $r$, implying that the area inside $R_c$ always satisfies $\omega < \omega_c$. We also check that all $R_c$ values in Table \ref{Table:DarkInterior} with $b = 0.01$ meet the dark energy conditions in Table \ref{Table:AnisotropicInteriorMatter} for $R \leq R_c$ (see Figure \ref{fig:dp} for a sample).

Similar to the standard interior case, we can obtain stable radiating gravastars with standard and repulsive phantom matter types for $k_1 = +1$. However, for $k_1 = -2$, only black hole structures are found. As a specific example, we provide the case $\omega = -0.5$, $b = 0.01$, and $\gamma = 0$ (dark energy interior with standard thin shell).

\begin{table}[H]
	\begin{tabular}{|c|c|c|c|c|c|c|c|}
		\hline
		\textbf{Case}&$\mathbf{\gamma}$&\multicolumn{2}{c|}{$\mathbf{k_1=0}$} & \multicolumn{2}{c|}{$\mathbf{k_1=+1}$} & \multicolumn{2}{c|}{$\mathbf{k_1=-2}$} \\
		\hline
		\multicolumn{2}{|c|}{}&$\mathbf{m_c}$ & $\mathbf{R_c}$ & $\mathbf{m_c}$ & $\mathbf{R_c}$ & $\mathbf{m_c}$  & $\mathbf{R_c}$ \\
		\hline
		C &-1 & 0.122662  &2.081189  &0.382801  &2.512076  & 1.117203 & 3.230482 \\
		
		& 0 & 0.266452 & 2.431494  & 0.468225 & 2.471987 &0.586041  & 1.316337 \\
		
		& 7/4 &0.803112 & 2.098185 &0.503750 & 1.015972 & -  & - \\
		
		& 3 & 0.510692 & 1.042769 &0.122662& 2.081189 &- &- \\
		\hline
		D &-1 & 0.110445 & 2.139964 & 0.363718 & 2.640231  & 1.141625 & 3.331433\\
		
		&0  & 0.248257  & 2.533278  & 0.450737 & 2.617435 & 0.587140 & 1.319180 \\
		
		& 7/4 & 0.809399 & 2.121009 & 0.503891 & 1.016243& -  & -\\
		
		& 3 &0.510931 & 1.043238 &  0.110445 & 2.139964 & - & -   \\
		\hline	
	\end{tabular}
	\caption{Comparison of critical mass $m_c$ and critical radius $R_c$ between the radiating case ($k_1=+1$, $k_1=-2$) and the non-radiating case ($k_1=0$) with repulsive phantom energy interior. Case C: $\omega=-0.5$ and $b=0.01$ and case D: $\omega=-0.7$ and $b=0.01$.}
	\label{Table:DarkInterior}
\end{table}

\begin{figure}[H]
	\centering
	\begin{subfigure}{0.48\textwidth}
		\includegraphics[width=\textwidth]{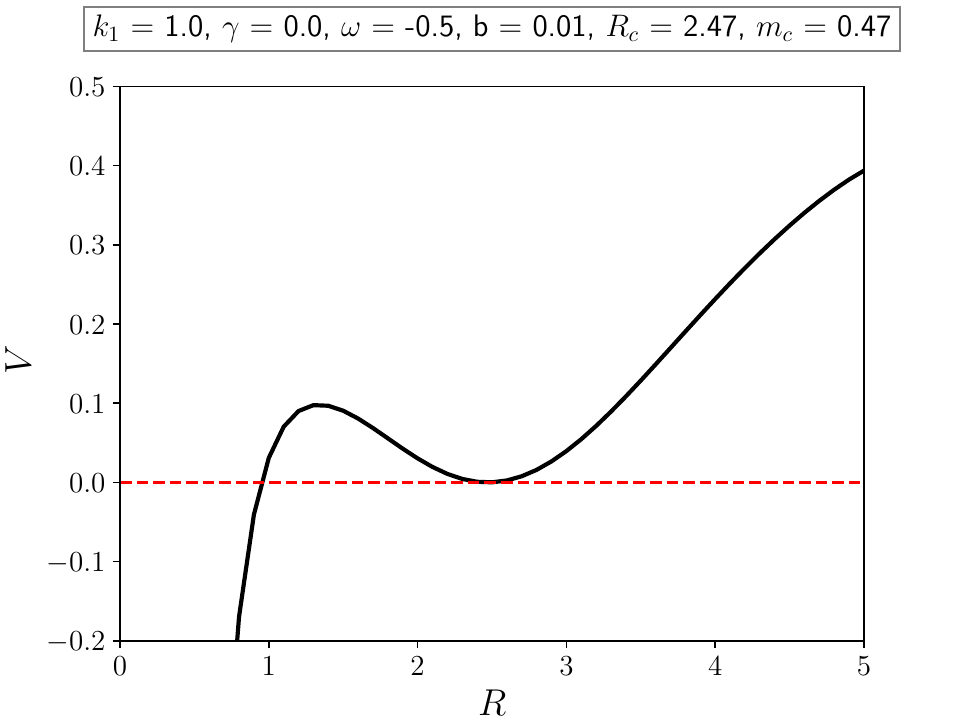}
		\caption{}
	\end{subfigure}
	\begin{subfigure}{0.48\textwidth}
		\includegraphics[width=\textwidth]{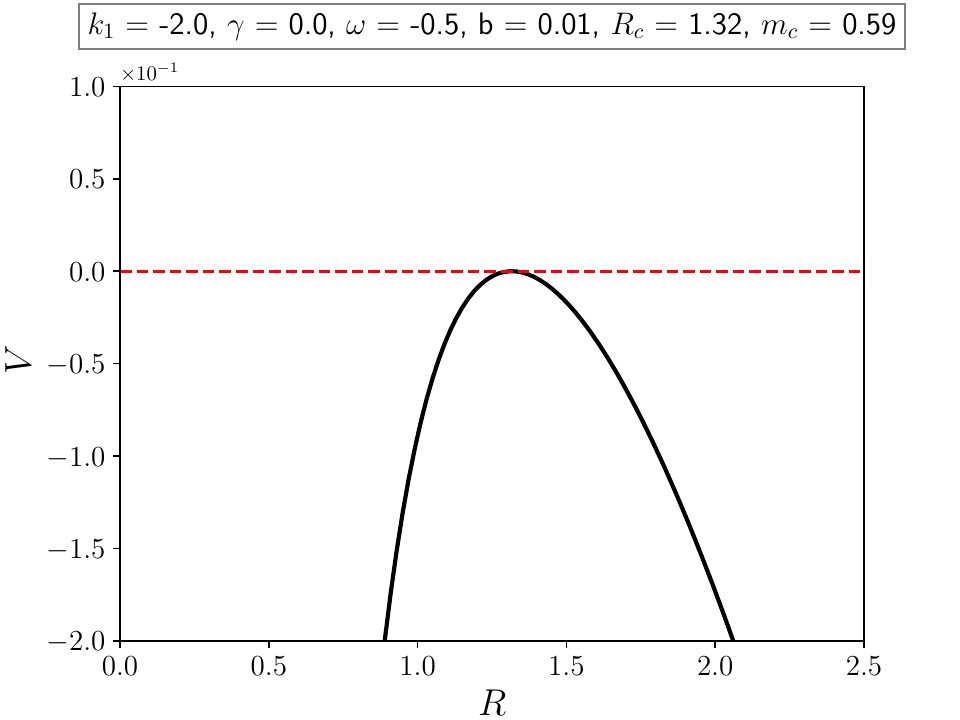}
		\caption{}
	\end{subfigure}
	\caption{The potential $V(R)$ for $\gamma=0$, $\omega=-0.5$, and $b=0.01$ with (a) $k_1=+1$ and (b) $k_1=-2$. We note that for $k_1=+1$ case, stable gravastar can be found, while for $k_1=-2$, only black hole is found.}
\end{figure}

\subsection{Repulsive Phantom Interior Case}

Repulsive phantom energy can be found at $\omega < -1$. In this case, for $k_1 = +1$, we can obtain both radiating gravastars and black hole structures with standard and repulsive phantom thin shells. 
However, for a dark energy shell, we only obtain black hole structures. Additionally, as with the other two types of thin shells, for $k_1 = -2$, we only obtain black hole structures.

\begin{figure}[H]
	\centering
	\includegraphics[width=0.6\linewidth]{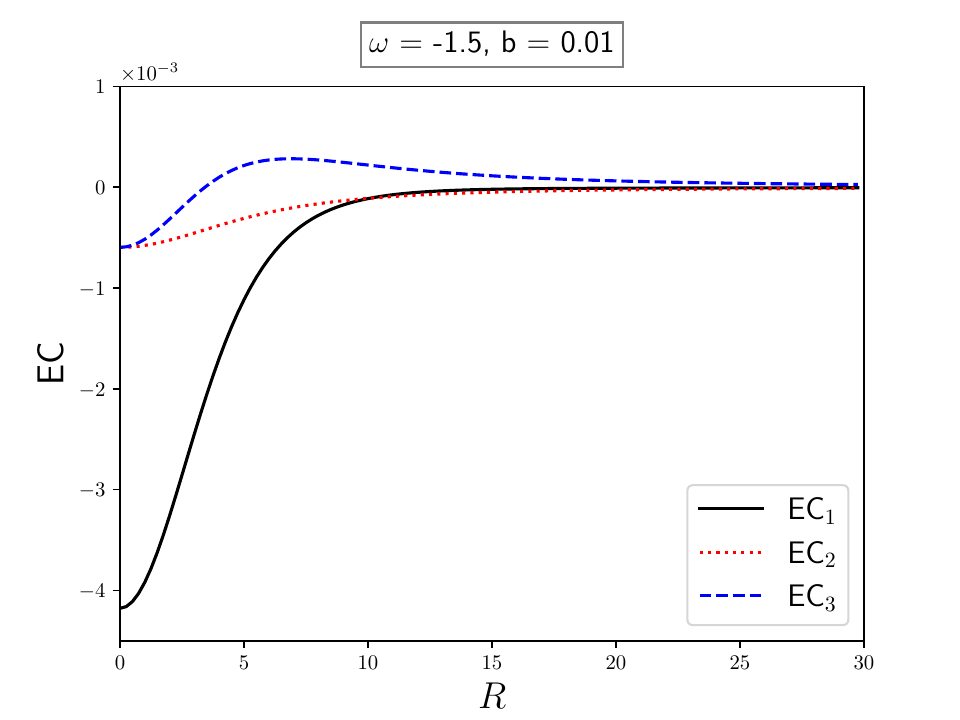}
	\caption{Energy condition $\text{EC}_1\equiv \rho+p_r+2p_t$, $\text{EC}_2\equiv \rho+p_r$ and $\text{EC}_3\equiv \rho+p_t$ for $\gamma=-1.5$ and $b=0.01$. We can see that $\text{EC}_1$ and $\text{EC}_2$ are negative and $\text{EC}_3$ is negative for small $R$ and positive for big $R$. This condition is repulsive phantom energy interior condition.}
	\label{fig:jp}
\end{figure}

We analyze two combinations of $\omega$ and $b$: $\omega = -1.25$ with $b = 0.01$ and $\omega = -1.5$ with $b = 0.01$. In both combinations, we clearly identify repulsive phantom energy. Our classification differs in this case, as the previous work by R. Chan et al. classified an interior with $\omega = -1.5$ and $b = 0.01$ as attractive phantom energy. To highlight this contrast, we show the energy conditions graph in Figure \ref{fig:jp} for $\omega = -1.5$ and $b = 0.01$. We also note here that these two cases fulfill negative gravity profile $(g(r)<0)$ condition which confirms that these cases are repulsive, not attractive. As a sample, we present graphs for the radiating case with interior parameters $\omega = -1.5$ and $b = 0.01$ and thin shell parameter $\gamma = 0$.

\begin{table}[H]
	\begin{tabular}{|c|c|c|c|c|c|c|c|}
		\hline
		\textbf{Case}&$\mathbf{\gamma}$&\multicolumn{2}{c|}{$\mathbf{k_1=0}$} & \multicolumn{2}{c|}{$\mathbf{k_1=+1}$} & \multicolumn{2}{c|}{$\mathbf{k_1=-2}$} \\
		\hline
		\multicolumn{2}{|c|}{}&$\mathbf{m_c}$ & $\mathbf{R_c}$ & $\mathbf{m_c}$ & $\mathbf{R_c}$ & $\mathbf{m_c}$  & $\mathbf{R_c}$ \\
		\hline
		E & -1 &0.505965   &1.024724  &0.531757  &1.151440   &  1.226145 & 3.703151 \\
		&  & 2.707314 & 35.645325 & 0.283417 &3.330909   &  & \\
		\hline
		& 0  &0.517835 & 1.079631  &0.5946233 & 1.234883  &0.590222  & 1.327173  \\
		& & 2.706589  &35.654932 & 0.372102 & 3.406399 &  & \\
		\hline
		& 7/4 & 0.828189& 2.190860 & 0.504279 & 1.016990 & -  & -\\
		\hline
		& 3 &0.511590 & 1.044534 & 0.505965   & 1.024724 & - & -   \\
		
		& & & & 2.707314  & 35.645325  &  &   \\
		\hline
		F & -1 & 0.505744  &1.024188  &0.530947  &1.148261   & 1.277690 &3.950008 \\
		&  & 0.163176 & 8.558249 & 0.109081 & 9.006250   &  & \\
		\hline
		& 0  &0.517369  & 1.078166  &0.545074  & 1.228637  &0.591651  & 1.330895  \\
		&  & 0.150217 &8.279282  &0.052395  &9.218837  &  & \\
		\hline
		& 7/4 & 0.837580 & 2.226766 & 0.504456 & 1.017330 & -  & -\\
		\hline
		& 3 &0.511891 & 1.045126 & 0.505744  & 1.024188 & - & -   \\
		
		& & & & 0.163176  & 8.558249 &  &   \\
		\hline
	\end{tabular}
	\caption{Comparison of critical mass $m_c$ and critical radius $R_c$ between the radiating case ($k_1=+1$, $k_1=-2$) and the non-radiating case ($k_1=0$) with repulsive phantom energy interior. Case E: $\omega=-1.25$ and $b=0.01$ and case F: $\omega=-1.5$ and $b=0.01$.}
	\label{Table:RepulsivePhantomInterior}
\end{table}

\begin{figure}[H]
	\centering
	\begin{subfigure}{0.48\textwidth}
		\includegraphics[width=\textwidth]{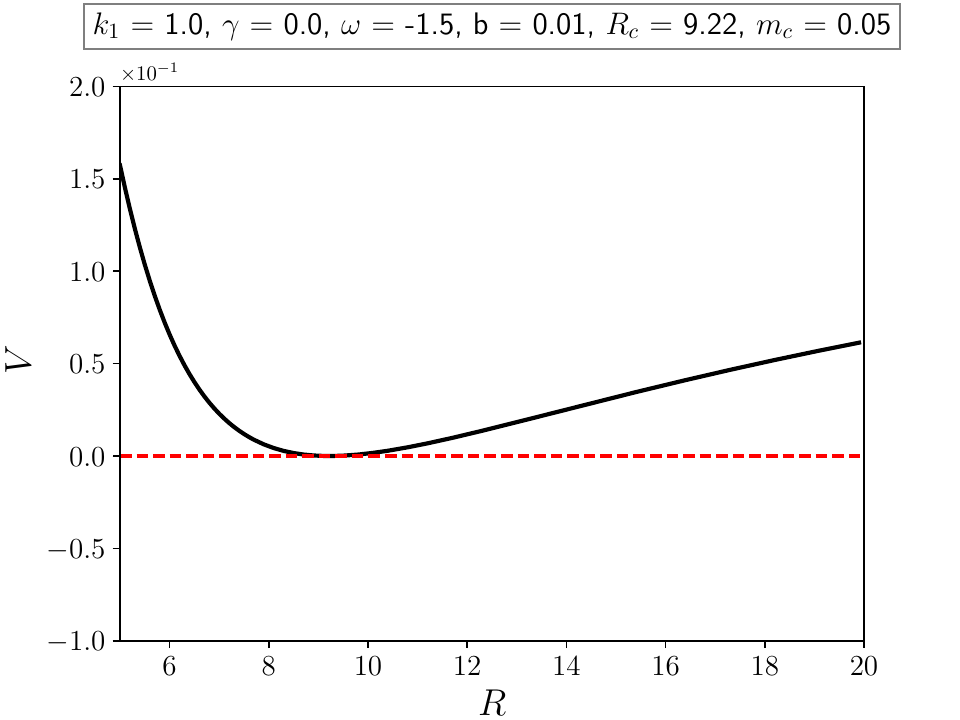}
		\caption{}
	\end{subfigure}
	\begin{subfigure}{0.48\textwidth}
		\includegraphics[width=\textwidth]{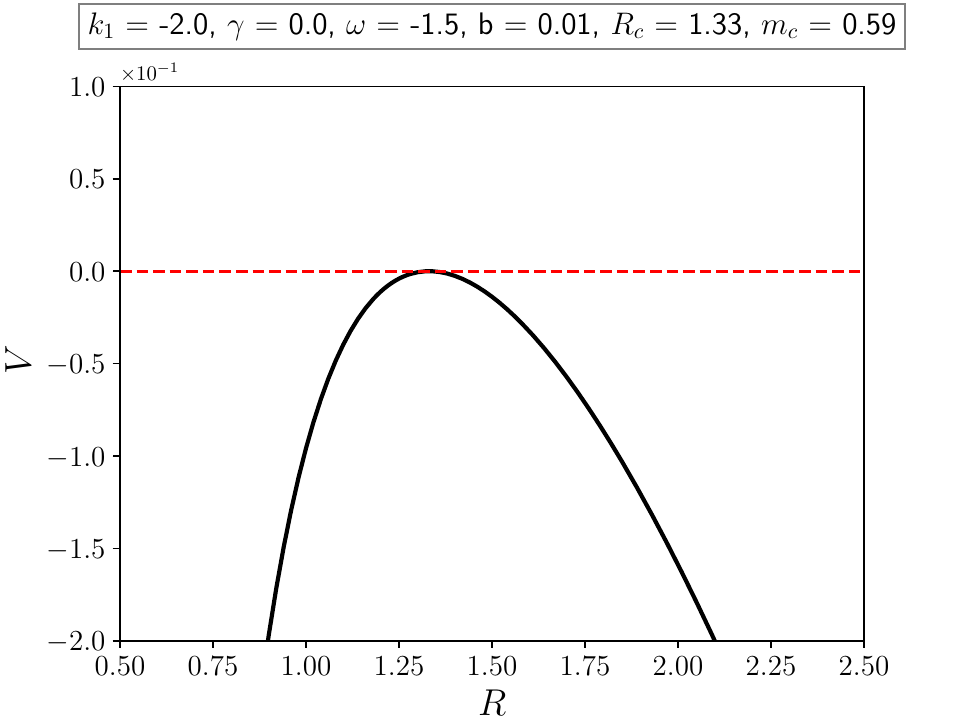}
		\caption{}
	\end{subfigure}
	\caption{The potential $V(R)$ for $\gamma=0$, $\omega=-1.5$, and $b=0.01$ with (a) $k_1=+1$  and (b) $k_1=-2$. We note that for $k_1=+1$ case, stable gravastar can be found, while for $k_1=-2$, only black hole is found.}
\end{figure}

\subsection{The Possible Scenarios and Characteristics}
The structures of cases A-F are summarized in Table \ref{Table:Scenarios}. From this table, it can be seen that radiating normal stars or gravastar structures can be found in cases with standard and repulsive phantom shells. The radiating gravastars (for the two chosen values of $k_1$) can only be obtained in the case of $k_1 = +1$. For the case $k_1 = -2$, only black hole structures are found. Therefore, the possible scenario for normal stars or gravastars with a standard shell is that the mass of the shell decreases because $k_1 (1-\gamma) > 0$. Meanwhile, for gravastars with a repulsive phantom shell, the scenario involves an increase in shell mass because $k_1 (1-\gamma) < 0$.

Additionally, from the data presented in Tables \ref{Table:StandardInterior}, \ref{Table:DarkInterior}, and \ref{Table:RepulsivePhantomInterior}, one important piece of information we can derive is that $m_c/R_c$ of black holes is always greater compared to gravastars or normal stars for each case. This indicates that the density of black holes is always greater compared to $m_c/R_c$ of gravastars and normal stars, at least for the same interior and shell matter types. This shows that gravastars have different characteristics (at least in terms of their compactness) than black holes. This point can strengthen the conclusion of Rocha, et al. in some of their papers \cite{rocha2008stable,rocha2008bounded}.

\begin{table}[H]
	\centering
	\begin{tabular}{|c|c|c|c|c|}
		\hline
		\textbf {Interior Energy} & \textbf{Shell Energy} & $\mathbf{k_1=0}$ & $\mathbf{k_1=+1}$ & $\mathbf{k_1=-2}$ \\
		\hline
		Standard		   & Standard          & Normal Star & Normal Star & Black Hole\\
		Standard		   & Dark             & Black Hole & Black Hole  & -\\
		Standard            & Repulsive Phantom & Black Hole& Normal Star & -\\
		
		Dark   & Standard          & Gravastar  & Gravastar & Black Hole\\
		Dark & Dark              & Black Hole & Black Hole   &- \\
		Dark &  Repulsive Phantom & Black Hole & Gravastar    & -\\
		Repulsive Phantom & Standard          & Gravastar/ & Gravastar/ & Black Hole\\
		&           & Black Hole  & Black Hole & \\
		Repulsive Phantom & Dark              & Black Hole & Black Hole  & - \\
		Repulsive Phantom &  Repulsive Phantom & Black Hole & Gravastar/ & - \\
		&   & & Black Hole &  \\
		\hline
	\end{tabular}
	\caption{Possible formation for various interior and thin shell matters. The case $k_1=0$ denotes non-radiating case and $k_1=+1$ and $k_1=-2$ denote radiating cases. The symbol "-" means none structure in that case.}
	\label{Table:Scenarios}
\end{table}


\section{Conclusion} 
\label{sec:conclusion}
	
In this paper, we present our study about radiating gravastars for both isotropic and anisotropic interior case by modifying the model of R. Chan et al. \cite{chan2011radiating}. For isotropic interior case, we primarily focus on investigating de Sitter spacetime interior with thin shell modification. In addition to this isotropic case, we examine the possibility to find gravastar formation with anti-de Sitter spacetime interior. For anisotropic interior, we only concentrate in exploring Tolman-Matese-Whitman (TMW) model. Our findings indicate that the potentials of both isotropic and anisotropic case depend significantly on the values of $\eta$, which is a function of $k_1$, $k_2$, and $\gamma$ for the isotropic case, and $k_1$ and $\gamma$ for the anisotropic.

In the study of isotropic interior, we consider changes in both mass and radius. We define $k_1$ as the constant for shell mass change and $k_2$ as the constant for radius change. For de Sitter interior case, we find that for $k_1 = k_2$, our results align with those of R. Chan et al. \cite{chan2011radiating} for all $\gamma$ values. When $k_1$ and $k_2$ are distinct, $\eta$ varies with each $\gamma$. Our results show that gravastar formation is possible only when $k_1 = +1$ for all $\gamma$. The scenario for radiating de Sitter gravastars with a normal matter thin shell involves decreasing mass and radius (for $\gamma = -1$) or a static radius (for $\gamma = 0$). For radiating de Sitter gravastars with dark energy thin shell, we find two possible scenarios: both mass and radius increase and both mass and radius decrease. While gravastars with repulsive phantom energy thin shells, the mass and radius both increase.

For anti-de Sitter interior case, we also obtain radiating gravastar formations for every case of thin shell type with specific values of $k_{1}$ and $k_{2}$. We find that, in general, the radiating scenarios of anti-de Sitter gravastars are different than de Sitter ones for the same type of thin shell. In addition, they are different in comparation between value of $R_c$ and $L_c$.

Our study of isotropic case generalizes previous studies on isotropic stable gravastars, especially de Sitter interior, with a three-layer model for both non-radiating and radiating case \cite{rocha2008bounded,rocha2008stable,chan2011radiating}. We formulated the potential, its first and second derivatives, and the mass as functions of $\eta$. Future work could focus on identifying the range or specific values of $\eta$ that ensure stability, and finding the exact solution for $\dot{m}\dot{v}^{3}$ without assumptions.

In the anisotropic section, we demonstrated several stable radiating gravastars with an anisotropic dark energy fluid interior and TMW mass function. We identified three possible interior types: standard, dark, and repulsive phantom, but not attractive phantom. Only the dark and repulsive phantom interiors can form gravastars. Our classification differs from that of R. Chan et al. \cite{chan2009stable} due to the different sign in the first term of $p_t$—we use a positive sign, while they used a negative one.

We also compared radiating and non-radiating cases of anisotropic gravastars. Our model shows that gravastars or normal stars (for a standard interior) can be found in stars with a standard shell, both radiating and non-radiating. If the shell is dark energy, only black holes are found. For a repulsive phantom shell, normal stars or gravastars are found in radiating cases, while non-radiating cases yield black holes. This suggests that gravastars are most commonly found with a standard shell type.

Lastly, our data show that the density of black holes is always greater than that of gravastars or normal stars, at least for similar types of interiors and thin shells. This supports the conclusion of R. Chan et al. that even if gravastars exist, they do not exclude the existence of black holes \cite{rocha2008bounded,rocha2008stable,chan2009stable}. Gravastars and black holes can thus be considered distinct objects.

\section{Acknowledgements}

The work of this paper is supported by RISET ITB 2024.

\printbibliography

\end{document}